\documentclass[a4paper,11pt]{article}
\usepackage{jheppub}

\usepackage{amsmath,amssymb,amsfonts,amsthm,amscd,bm}
\usepackage{graphicx}
\usepackage{float}
\usepackage{longtable}
\usepackage{booktabs}
\usepackage{enumitem}

\restylefloat{table}

\usepackage{caption,subcaption}

\usepackage{tikz}

\def\fg{\mathfrak{g}}

\def\bbC{\mathbb{C}}
\def\bbZ{\mathbb{Z}}

\def\CN{{\cal N}}

\def\tr{\mathop{\rm tr}}

\def\beq#1\eeq{\begin{align}#1\end{align}}

\title{On the Representation Theory of Non-Admissible $W$-Algebras: Part I}

\author{Dan Xie}
\emailAdd{danxie@mail.tsinghua.edu.cn}

\affiliation{Department of Mathematics, Tsinghua University, Beijing, 100084, China}

\abstract{Motivated by the mirror symmetry for circle compactified 4d $\mathcal{N}=2$ theories, we propose a geometric framework for studying the representation theory of non-admissible $W$-algebras $W^{k}(\mathfrak g,f)$ at levels $k=-h^\vee+\frac{1}{n}\frac{m}{u}$, using the geometry of generalized affine Springer fibers $Sp_{\nu}(\tilde{\mathfrak g},f)$ with slope $\nu=u/m$.  The central proposal is that each non-empty $\bbC^*$-fixed locus, labeled by a double coset $\tilde w\in W_\nu\backslash\tilde{W}/W_f$, gives rise to simple modules whose highest weight is determined by the map $\tilde w\mapsto\tilde w(k\Lambda_0+\tilde\rho)-\tilde\rho$, while the dimension of the fixed variety encodes additional non-semisimple structure (logarithmic modules). We verify this correspondence in numerous examples, including $D_4$, $E_6$, $E_8$, and twisted theories of type $^3D_4$ and $^2A_3$, where our geometric counting reproduces known results for both admissible and non-admissible $W$-algebras.}

\begin{document}
\maketitle
\flushbottom

\section{Introduction}
Vertex operator algebras (VOAs) arise naturally from the chiral sector of two-dimensional conformal field theories \cite{belavin1984infinite}.
One of the central problems in the study of VOAs is to classify all VOA modules.
Once the modules are understood, one can study modular transformations of characters and extract operator product expansion (OPE) data via the Verlinde formula; this framework is especially effective for rational VOAs \cite{moore1989classical}.

The classification of VOA modules is difficult in general, and several complementary tools are available.
One approach uses Zhu's algebra \cite{zhu1996modular}, which reduces representation-theoretic questions for a VOA to questions in associative algebras.
Another uses Drinfeld--Sokolov (DS) reduction \cite{Frenkel:1992ju}, which relates representations of $W$-algebras to representations of affine Kac--Moody algebras, which are easier to study.

Most existing results concern rational VOAs, whose module categories are semisimple and have only finitely many simple objects; Virasoro and $W$ minimal models are standard examples.
It is important to extend this analysis to VOAs with non-semisimple representation categories.
A natural and tractable class is given by $C_2$-cofinite VOAs \cite{zhu1996modular,miyamoto2004modular}, where logarithmic modules may appear \cite{flohr1996modular,feigin2006logarithmic} yet the algebra retains key finiteness properties.
Many structural features survive in this setting, including modular properties of generalized characters and Verlinde-type formulas, but explicit representation-theoretic results are still limited.

In this paper, we study non-admissible $W$-algebras of the form
\begin{equation*}
W^{k}(\mathfrak g,f),\qquad k=-h^\vee+\frac{1}{n}\frac{m}{u},
\end{equation*}
where $f$ is a nilpotent orbit and $u$ is a positive integer coprime to $m$ (see \cite{Frenkel:1992ju,deBoer:1993iz,kac2003quantum} for the definition of $W$-algebras).
The allowed values of $m$ for each Lie algebra $\mathfrak g$ are listed in Table~\ref{table:ellnumber}, and $n$ is the lacing number (see Table~\ref{table:lacing}).
For suitable data $(\mathfrak g, k, f)$, many of these $W$-algebras are expected to be $C_2$-cofinite \cite{Xie:2019vzr}.

When $m=h^\vee$, one obtains (boundary) admissible $W$-algebras \cite{kac2017remark}, whose representation theory can often be analyzed using DS reduction \cite{Frenkel:1992ju,Arakawa2012Rationality}.
When $m\neq h^\vee$, the $W$-algebra is non-admissible, and much less is known from a purely VOA perspective (see \cite{Arakawa_2016} for some examples).

Our main tool is the 4d mirror symmetry viewpoint \cite{shan2026mirror,shan2024modularity}, which links VOA representation theory to Hitchin moduli spaces and generalized affine Springer fibers.
It was shown in \cite{shan2026mirror} that the representation theory of the $W$-algebra is recovered elegantly from the geometry of the generalized Springer fiber studied in \cite{goresky2003purity}.
In our setting, the relevant geometry is controlled by the slope $\nu=\frac{u}{m}$ and by the parabolic subalgebra associated with the nilpotent orbit $f$, and so the fiber is denoted by $Sp_{\nu}(\tilde{\mathfrak g}, f)$.
For $m=h^\vee$, known results support a correspondence between fixed points of affine Springer fibers under a $\bbC^*$ action \footnote{This $\bbC^*$ action corresponds to the $U(1)_R$ symmetry of a four-dimensional $\mathcal{N}=2$ SCFT.}
and simple VOA modules \cite{shan2026mirror,shan2024modularity}.

The main purpose of this paper is to extend this correspondence to the case $m\neq h^\vee$. On the $W$-algebra side, logarithmic modules are expected to appear; on the affine Springer fiber side, the fixed loci can have positive dimension.
This motivates the following geometric representation-theoretic proposal:
\begin{enumerate}
\item Fixed loci of $Sp_{\nu}(\tilde{\mathfrak g}, f)$ are labeled by affine Weyl-group elements $\tilde w$ \cite{goresky2003purity,oblomkov2016geometric,varagnolo2009finite}; each $\tilde{w}$ determines one or more simple modules and their conformal weights (see formula~\ref{eq:highest-weight-map}).
\item The dimension of the fixed variety $S^{\tilde w}_0$ controls additional non-semisimple structure, including logarithmic modules and generalized characters (see \cite{pan2024mirror} for a similar identification for class-$A_1$ theories).
\end{enumerate}

The set of relevant $\tilde w$ and the dimensions of the fixed varieties $S^{\tilde w}_0$ can be computed explicitly, as summarized in Section~\ref{subsec:recap}.
Combining this with the relation between affine Springer cohomology and DAHA actions suggests a geometric route to modular data of the VOA \cite{shan2024modularity}.
Our main conjecture is that, for $C_2$-cofinite non-admissible $W$-algebras, this construction captures the full representation-theoretic and modular content.
In this paper, we primarily examine the space of possible representations and leave the modular behavior for later study.

Since very little is known about non-admissible $W$-algebras, we employ the following checks: for special choices of the slope and the nilpotent orbit $f$, the 4d theory predicts that such a $W$-algebra is isomorphic to a known admissible $W$-algebra; we then verify that our computations reproduce the known results.

The paper is organized as follows.
Section~2 reviews the 4d mirror symmetry setup and its relation to generalized affine Springer fibers.
Section~3 summarizes the geometric counting of fixed loci and their dimensions.
Section~4 reviews the $W$-algebra data relevant for comparison with geometry.
Section~5 discusses the regular and subregular slopes.
Section~6 gives detailed computations for untwisted theories focusing on examples of types $D_4$, $E_6$, and $E_8$.
Section~7 discusses the twisted theories.
Section~8 concludes. Conventions and additional computational examples are collected in the appendix.

\begin{table}[htp]
    \begin{center}
    \begin{tabular}{|c|c|c|c|c|}\hline
    $\mathfrak g$ & $\dim\mathfrak g$ & $h$ & $h^\vee$ & $n$ \\ \hline
    $A_{N-1}$ & $N^2-1$ & $N$ & $N$ & $1$ \\ \hline
    $B_N$ & $(2N+1)N$ & $2N$ & $2N-1$ & $2$ \\ \hline
    $C_N^{(1)}$ & $(2N+1)N$ & $2N$ & $N+1$ & $4$ \\ \hline
    $C_N^{(2)}$ & $(2N+1)N$ & $2N$ & $N+1$ & $2$ \\ \hline
    $D_N$ & $N(2N-1)$ & $2N-2$ & $2N-2$ & $1$ \\ \hline
    $E_6$ & $78$ & $12$ & $12$ & $1$ \\ \hline
    $E_7$ & $133$ & $18$ & $18$ & $1$ \\ \hline
    $E_8$ & $248$ & $30$ & $30$ & $1$ \\ \hline
    $F_4$ & $52$ & $12$ & $9$ & $2$ \\ \hline
    $G_2$ & $14$ & $6$ & $4$ & $3$ \\ \hline
    \end{tabular}
    \end{center}
    \caption{Lie algebra data: $h$ is the Coxeter number, $h^\vee$ is the dual Coxeter number, and $n$ is the lacing number. For the symplectic algebras $C_N$, two normalizations are commonly used; $C_N^{(2)}$ ($n=2$) is the standard lacing convention used in the $W$-algebra literature.}
    \label{table:lacing}
\end{table}

\begin{table}[htp]
    \begin{center}
    \begin{tabular}{|c|c|} \hline
    $\mathfrak{j}$& Elliptic number $m$ \\ \hline
    $A_{n}$& $n+1$ \\  \hline
    $D_n$&    $m~\text{even} ,~ \frac{2n-2}{m} ~\text{odd}$  \\  \hline
    ~& $m~\text{even} ,~ \frac{2n}{m} ~\text{even}$ \\ \hline
    $E_6$& $12,9,6,3$ \\ \hline
    $E_7$& $18,14,6,2$ \\ \hline
    $E_8$& $2, 3, 5, 6, 10, 15, 30$ \\ \hline
    ~&$4, 8, 12, 24$ \\ \hline
    ~&$20$ \\ \hline
    \end{tabular}
    \quad
    \begin{tabular}{|c|c|} \hline
    $\mathfrak{j},o$& Elliptic number $m$ \\ \hline
    $A_{2n},\bbZ_2$& $m=2r,~r~\text{odd} ,~ \frac{2n+1}{r} ~\text{odd}$ \\  \hline
    ~&$m=2r,~r~\text{odd} ,~ \frac{2n}{r} ~\text{even}$ \\ \hline
    $A_{2n-1},\bbZ_2$&    $m=2r,~r~\text{odd} ,~ \frac{2n-1}{r} ~\text{odd}$  \\  \hline
    ~& $m=2r,~r~\text{odd} ,~ \frac{2n}{r} ~\text{even}$ \\ \hline
    $D_n,\bbZ_2$& $m~\text{even} ,~ \frac{2n}{m} ~\text{odd}$  \\ \hline
    ~&$m~\text{even} ,~ \frac{2n-2}{m} ~\text{even}$  \\ \hline
    $D_4,\bbZ_3$& $12,6,3$ \\ \hline
    $E_6,\bbZ_2$& $18,12,6,4,2$ \\ \hline
    \end{tabular}
    \end{center}
    \caption{List of elliptic numbers $m$.}
    \label{table:ellnumber}
\end{table}

\newpage
\section{4d mirror symmetry}
We review the 4d mirror symmetry framework described in \cite{shan2026mirror} (see earlier work in \cite{Fredrickson:2017yka,Fredrickson:2017jcf}).
The relation between $W$-algebras and Hitchin moduli spaces, with affine Springer fibers appearing as zero fibers, is naturally realized in four-dimensional $\CN=2$ theories.
The $W$-algebra captures the Schur sector (containing the Higgs branch as a subsector), while Hitchin geometry captures Coulomb-branch data.
From this perspective, 4d mirror symmetry links Schur-sector observables to Coulomb-branch geometry of a single 4d $\mathcal{N}=2$ theory, extending the familiar 3d mirror correspondence,
which relates the Higgs branch and the Coulomb branch of a 3d $\mathcal{N}=4$ theory.

\begin{figure}
    \begin{center}

    \tikzset{every picture/.style={line width=0.75pt}} 

    \begin{tikzpicture}[x=0.55pt,y=0.55pt,yscale=-1,xscale=1]

    \draw   (352,224.75) .. controls (352,143.7) and (417.7,78) .. (498.75,78) .. controls (579.8,78) and (645.5,143.7) .. (645.5,224.75) .. controls (645.5,305.8) and (579.8,371.5) .. (498.75,371.5) .. controls (417.7,371.5) and (352,305.8) .. (352,224.75) -- cycle ;
    \draw   (492,125) -- (500.5,125) -- (500.5,133.5) -- (492,133.5) -- cycle ;
    \draw    (489.5,292.96) -- (501.5,305.96) ;
    \draw    (488.5,305.96) -- (501.5,291.96) ;

    \draw (462,123.4) node [anchor=north west][inner sep=0.75pt]    {$\Phi $};
    \draw (466,283.4) node [anchor=north west][inner sep=0.75pt]    {$f$};
    \draw (490,41.4) node [anchor=north west][inner sep=0.75pt]    {$(\mathfrak{j},o)$};

    \end{tikzpicture}

\caption{A sphere with one irregular singularity $\Phi$ and one regular singularity labeled by $(\mathfrak j,o)$ (or nilpotent orbit $f$).}
\label{fig:setup_sphere_irregular_regular}
\end{center}
\end{figure}
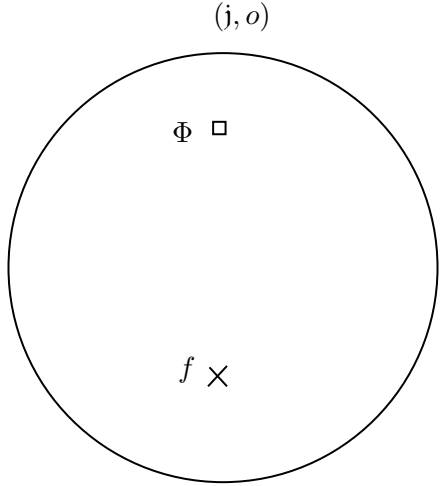

A large class of examples comes from compactifying the 6d $(2,0)$ theory of type $\mathfrak j=ADE$ on a sphere with one irregular singularity $\Phi$ and one regular singularity labeled by a nilpotent orbit $f$, see Figure~\ref{fig:setup_sphere_irregular_regular}.
Near the irregular singularity, the Higgs field behaves as
\begin{equation*}
\Phi=\frac{T}{z^{2+\frac{k}{m}}}+\cdots,
\end{equation*}
where $T$ is a regular semisimple element of $\mathfrak{g}$.
The allowed values of $m$ form a finite set whose maximal element is the Coxeter number (see \cite{Xie:2012hs,Wang:2015mra,Wang:2018gvb}), and we assume $k>-m$.

After the coordinate change $z'=1/z$, and using the fact that $\Phi$ is a section of the canonical bundle of the sphere, we obtain
\begin{equation*}
\Phi=Tz^{\frac{k}{m}}+\cdots.
\end{equation*}
To match affine Springer conventions, we consider
\begin{equation*}
\Psi(z)=z\Phi(z)=Tz^{\frac{u}{m}},\qquad u>0,\qquad u=k+m.
\end{equation*}
This defines an element $v$ in the affine Lie algebra.
The dense open part of the zero Hitchin fiber can be written as
\begin{equation*}
H_0=\left\{g\in G(z,z^{-1})/G(z)\,\middle|\,g^{-1}v\in \tilde{\mathfrak{n}}_f\right\},
\end{equation*}
where $\tilde{\mathfrak{n}}_f$ is a subalgebra of the affine Lie algebra.
This is precisely the generalized affine Springer fiber studied in \cite{goresky2003purity}.

The 4d theory has several protected sectors, in particular the Coulomb branch and the Schur sector.
The Coulomb branch information is encoded by the Hitchin spectral curve, while the Schur sector is described in \cite{Xie:2016evu,Song:2017oew,Wang:2018gvb} by a two-dimensional VOA.
When the irregular singularity has no mass parameters (for the allowed values of $m$ in Table~\ref{table:ellnumber}), the associated 2d VOA for the Schur sector is
\begin{equation*}
W^{k}(\mathfrak{j},f),\qquad k=-h^\vee+\frac{m}{u}.
\end{equation*}
On the affine Springer side, the no-mass-parameter condition implies that $v$ is regular elliptic and homogeneous, a case studied in detail in \cite{oblomkov2016geometric}.
There is a $\bbC^*$ action on the corresponding affine Springer fiber; physically, this reflects superconformal symmetry of the 4d $\mathcal{N}=2$ theory.

For non-simply-laced $\mathfrak g^\vee$, one must include an outer-automorphism twist \cite{Wang:2018gvb} of an ADE algebra $\mathfrak j$.
The invariant algebra is denoted by $\mathfrak g$, which is Langlands dual to $\mathfrak g^\vee$.
Then the irregular singularity defines an element in a twisted affine Lie algebra, while the Schur sector is described by the untwisted affine Lie algebra of type $\mathfrak g^\vee$:
\begin{equation*}
W^{k}(\mathfrak{g}^\vee,f),\qquad k=-h^\vee+\frac{1}{n}\frac{m}{u}.
\end{equation*}
Here $n$ is the lacing number.
The generalized affine Springer fiber is described by the twisted affine algebra of dual type to $\mathfrak g^\vee$ (the invariant part of the twisted affine Lie algebra is $\mathfrak g$, reflecting Langlands duality between untwisted and twisted affine systems).

Therefore the 6d construction directly relates two classical objects: $W$-algebras and generalized affine Springer fibers.
A key relation from \cite{shan2026mirror} is that representation-theoretic data of the $W$-algebra can be read from affine Springer cohomology.
Because the cohomology localizes to fixed loci of the $\bbC^*$ action, the problem reduces to studying fixed varieties labeled by affine Weyl-group elements $\tilde w$.
The corresponding highest weights on the $W$-algebra side are obtained from the simple map:
\begin{equation}
\boxed{\tilde{w} \to \tilde{w}(k\Lambda_0+\tilde{\rho})-\tilde{\rho}}
\label{eq:highest-weight-map}
\end{equation}
where $\Lambda_0$ is the zeroth affine fundamental weight and $\tilde\rho$ is the affine Weyl vector.
This correspondence has been checked in boundary-admissible examples \cite{shan2026mirror}; here we extend it to non-admissible cases. One must be careful about normalization in the twisted theories; see Section~\ref{section:twisted}.

The main new feature in the non-admissible case is that fixed varieties can have positive dimension.
On the VOA side, this is expected to correspond to the appearance of logarithmic modules.
For instance, $V_{-2}(D_4)$ has logarithmic modules, and the corresponding affine Springer fiber contains a one-dimensional fixed variety.
This motivates the proposal:
\begin{quote}
For a higher-dimensional fixed variety labeled by $\tilde w$, one obtains one irreducible highest-weight module (via the map above) together with additional logarithmic modules whose characteristic conformal dimension is that of the corresponding simple module.
\end{quote}

The case $f=\text{regular}$ is especially interesting: the corresponding principal $W$-algebra is lisse (hence $C_2$-cofinite) \cite{Xie:2019vzr}, and in this case our construction may describe the full set of simple and logarithmic modules.

In this paper we focus on establishing the correspondence between the fixed varieties of the affine Springer fiber and the modules of the $W$-algebra; a detailed study of modular transformations is deferred to Part II.

We also use the standard 4d/2d dictionary \cite{Beem:2013sza,Beem:2017ooy}:
\begin{enumerate}
\item The central charges satisfy $c_{2d}=-12 c_{4d}$ and $k_{2d}=-k_{4d}$.
\item The Higgs branch of the 4d theory is identified with the associated variety of the 2d theory.
\end{enumerate}

\section{Generalized affine Springer fibers}

As discussed in the previous section, the representation theory of non-admissible $W$-algebras is closely related to the topology of generalized affine Springer fibers,
labeled by the data $(\mathfrak{g}, o, \nu, f)$, where $\mathfrak{g}$ is the ADE Lie algebra, $o$ is the choice of outer automorphism, $\nu=\frac{u}{m}$ denotes the type of irregular singularity in the Hitchin context and
is called the slope in the context of affine Springer fibers, and $f$ denotes the regular singularity in the Hitchin context, associated with a nilpotent orbit and the corresponding parabolic subalgebra $P$.
In this section we recall the basic definitions of the generalized affine Springer fiber associated with the data $(\mathfrak{g}, o, \nu, f)$, and the counting formulas for fixed loci and their dimensions \cite{goresky2003purity}.

Let $\mathfrak g$ be a simple Lie algebra with Cartan subalgebra $\mathfrak h$.
Its (loop) affine Lie algebra is
\begin{equation*}
\tilde{\mathfrak g}=\mathfrak g\otimes\bbC[\epsilon,\epsilon^{-1}].
\end{equation*}

For $x\in\mathfrak h$ and $r\in\mathbb{Q}$, define the subspace
\begin{equation*}
\mathbf V_x(r)=\bigoplus_{\alpha(x)+m=r} \mathfrak g_\alpha\,\epsilon^m,
\end{equation*}
where $\alpha$ runs over roots (and $\mathfrak g_\alpha$ is the corresponding root space), with pairing induced by the Killing form.
We also define the filtered subspace
\begin{equation*}
\mathbf V_{x,r}=\prod_{r'\ge r}\mathbf V_x(r').
\end{equation*}

Fix $y\in\mathfrak h$ and let $v\in\tilde{\mathfrak g}$ and $s\in\mathbb{Q}$.
The generalized affine Springer fiber is
\begin{equation*}
\boxed{F_y(t,v)=\{g\in G/G_y\mid g^{-1}v\in\mathbf V_{y,t}\}},
\end{equation*}
with $v\in\mathbf V_{x,s}$.
Here $G_y$ has Lie algebra
\begin{equation*}
\mathfrak g_y=\bigoplus_{\alpha(y)+m\ge 0}\mathfrak g_\alpha\,\epsilon^m.
\end{equation*}

In our setup,
\begin{equation*}
\boxed{y=c\rho_P,\qquad s=\frac{u}{m},\qquad x=\frac{u}{m}\rho,\qquad t=\varepsilon},
\end{equation*}
where $\rho$ is the Weyl vector, $\rho_P$ determines a standard parabolic $P$, $(u,m)=1$, and $c,\varepsilon>0$ are sufficiently small.
The choice $t=\varepsilon>0$ is slightly different from the standard $t=0$ convention.
Indeed, for small $\varepsilon>0$, the condition $(\alpha,y)+m\ge t$ for the space $V_{y,t}$ is equivalent to $(\alpha,y)+m>0$.
Thus we keep only the nilpotent part of the subalgebra defined by $y$ in the definition of $V_{y,t}$ (whereas $t=0$ keeps the full parahoric piece).

Geometrically, this affine Springer fiber describes the locus in the Hitchin system with fixed local boundary data and fixed spectral curve at the origin of the Hitchin fibration (the zero Hitchin fiber).

We now describe the parabolic data.
Let $\Delta$ be the root system of $\mathfrak g$ and $\Delta_+$ the positive roots.
For $\Delta_P\subset\Delta_+$, define
\begin{equation*}
\mathfrak g_P=
\left(\mathfrak h\oplus\bigoplus_{\alpha\in\Delta_P}(\mathfrak g_\alpha\oplus\mathfrak g_{-\alpha})\right)
\oplus\mathfrak n_P,
\end{equation*}
with nilpotent radical
\begin{equation*}
\mathfrak n_P=\bigoplus_{\alpha\in\Delta_+\setminus\Delta_P}\mathfrak g_\alpha.
\end{equation*}

\textbf{Example.}
For $\mathfrak{sl}_n$, parabolic subalgebras are classified by partitions
$f=[n_1,n_2,\ldots\allowbreak,n_s]$. In this notation, the corresponding set of positive roots is
\begin{align*}
\Delta_P
&=\{\alpha_1,\ldots,\alpha_{n_1-1}\}\\
&\qquad\cup\{\alpha_{n_1+1},\ldots,\alpha_{n_1+n_2-1}\}\\
&\qquad\cup\cdots .
\end{align*}
Concretely:
\begin{itemize}
\item $[n]$ (equivalently $\Delta_P=\Delta_+$) gives the full Lie algebra;
\item $[1,\ldots,1]$ (equivalently $\Delta_P=\emptyset$) gives the Borel subalgebra.
\end{itemize}
These partitions coincide with the partitions $f$ appearing in the corresponding $W$-algebra.
The subgroup $W_P$ is generated by simple reflections corresponding to roots in $\Delta_P$.

The linear map $\rho_P$ on simple roots is defined as
\begin{equation*}
\rho_P(\alpha)=
\begin{cases}
0,& \alpha\in\Delta_P,\\
1,& \alpha\notin\Delta_P.
\end{cases}
\end{equation*}
Using $\rho_P$, we associate a generalized affine Springer fiber for the Hitchin system defined in the previous section.

We now recall the computation of the cohomology of the generalized affine Springer fiber \cite{goresky2003purity}.
Because there is a $\bbC^*$ action on generalized affine Springer fibers, their (equivariant) cohomology can be computed from fixed loci \cite{goresky2003purity}.
Combinatorially, one studies affine Schubert cells labeled by double cosets
$W_\nu\backslash\tilde W/W_P$, with slope $\nu=\frac{u}{m}$, and studies the intersection of each cell with the generalized affine Springer fiber.
The group $W_\nu$ is generated by affine reflections in roots lying on
\begin{equation*}
\boxed{L_{u/m}:\qquad (\alpha,x)+k=0\quad\Longleftrightarrow\quad u(\alpha,\rho)+mk=0}.
\end{equation*}
To determine non-empty cells, one also needs
\begin{equation}
\boxed{S_{u/m}:\qquad (\alpha,x)+k=\frac{u}{m}\quad\Longleftrightarrow\quad u(\alpha,\rho)+mk=u}.
\end{equation}
The affine reflections generated by $L_{u/m}$ naturally act on $S_{u/m}$.

Each cell $S_{\tilde w}$ of the generalized affine Springer fiber is an iterated affine bundle over a Hessenberg variety \cite{goresky2003purity}:
\begin{equation*}
S_0^{\tilde w}=\{g\in H_0/P\mid g^{-1}v\in F_{\tilde y}^tV_x(s)\},
\end{equation*}
where
\[
H_0=G_x(0),\qquad P=F_{\tilde y}^{0}G_x(0),
\]
and
\begin{equation*}
F_{\tilde y}^{t}V_x(s)=\operatorname{im}\!\left[V_{\tilde y,t}\cap V_{x,s}\to V_x(s)\right].
\end{equation*}
For $\tilde w=st_q$, our convention is $\tilde y={\tilde {w}}^{-1}(y)=s^{-1}(y)-q$.  We can identify the space: $H_0$ is the space $L_{u/m}$,
and $V_x(s)$ is the space $S_{u/m}$.

The non-emptiness condition for $S_0^{\tilde w}$ is computed as follows: first consider the cohomology of the flag manifold $H^*(G/P, Q_l)$, where $P$ is the parabolic subgroup determined by $\tilde{w}$. The cohomology group
is the symmetric algebra $S^{\bullet}$ on the coweight space $X^*(A)\oplus Q_l$ modulo the ideal $I$ generated by the Weyl group invariant elements that are homogeneous of strictly positive degree.
Whether $S_0^{\tilde w}$ is non-empty is determined as follows: let $m$ be the dimension of $V/F_y^t V$, where $V=S_{u/m}$, and compute the characters $\lambda_1,\ldots, \lambda_m$
for $A$ on the $B$-module $V/F_{\tilde{y}}^t V$. Then $S_0^{\tilde w}$ is empty if and only if $\lambda_1\cdots\lambda_m$ lies in the ideal $I$.

There are two simpler necessary conditions for non-emptiness of a cell. The first is the following \cite{oblomkov2016geometric,varagnolo2009finite}:
\begin{equation}
\tilde w^{-1}(\Delta_0)\text{ lies in a bounded region of the hyperplane arrangement determined by }S_{u/m}.
\label{eq:cell-nonempty}
\end{equation}
Here $\Delta_0$ is the fundamental alcove.

The second condition is that the Hessenberg variety has non-negative dimension.
\paragraph{Dimension of the Hessenberg variety $S_0^{\tilde w}$}
For $\tilde w=st_q$ satisfying \eqref{eq:cell-nonempty}, its dimension is given by the following formula:
\begin{equation}
\dim\!\left(S_0^{\tilde w}\right)=
\dim\!\left(\frac{V_x(0)}{V_x(0)\cap V_{\tilde y,0}}\right)
-\dim\!\left(\frac{V_x(s)}{V_x(s)\cap V_{\tilde y,t}}\right).
\label{eq:hessenberg-dim}
\end{equation}
The denominator of the first term counts affine roots $(\alpha,k)$ satisfying
\begin{equation}
(\alpha,x)+k=0,\qquad (-q,\alpha)+k+c\bigl(\rho_P,s(\alpha)\bigr)\geq 0,
\label{equ1}
\end{equation}
and the denominator of the second term becomes
\begin{equation}
(\alpha,x)+k=s,\qquad (-q,\alpha)+k+c\bigl(\rho_P,s(\alpha)\bigr)\geq t.
\label{equ2}
\end{equation}
Here the condition on an affine root $(\alpha,k)$ involving $\tilde{y}$ is
\begin{equation*}
    \alpha(\tilde{y})+k=(\alpha, s^{-1}(c\rho_P)-q)+k=(-q,\alpha)+c(s(\alpha), \rho_P)+k.
\end{equation*}

\textbf{Remark}: It is possible that the above two necessary conditions are satisfied, but the Hessenberg variety is empty.
One needs to use the definition of the Hessenberg variety to determine whether the Hessenberg variety is empty.

\paragraph{Dimension of the cell $S^{\tilde w}$}
Similarly, the dimension of the cell labeled by $\tilde{w}$ is given by the formula:
\begin{equation*}
\dim\!\left(S^{\tilde w}\right)=
\dim\!\left(\frac{V_{x,0}}{V_{x,0}\cap V_{\tilde y,0}}\right)
-\dim\!\left(\frac{V_{x,s}}{V_{x,s}\cap V_{\tilde y,t}}\right).
\end{equation*}
The denominator of the first term counts affine roots $(\alpha,k)$ satisfying
\begin{equation*}
(\alpha,x)+k\ge 0,\qquad (-q,\alpha)+k+c\bigl(\rho_P,s(\alpha)\bigr)\ge 0,
\end{equation*}
while the denominator of the second term counts those satisfying
\begin{equation*}
(\alpha,x)+k\ge s,\qquad (-q,\alpha)+k+c\bigl(\rho_P,s(\alpha)\bigr)\ge t.
\end{equation*}

\subsection{Two important parabolics}
For $f=0$ (the regular parabolic, i.e., the Borel subalgebra $B$, with $\rho_P(\alpha)=1$ for all positive roots), the inequalities
in formula~\ref{eq:hessenberg-dim} become
\begin{equation*}
(-q,\alpha)+k+c\bigl(\rho_P,s(\alpha)\bigr)\ge 0,
\qquad
(-q,\alpha)+k+c\bigl(\rho_P,s(\alpha)\bigr)\ge t
\end{equation*}
and both of them simplify to
\begin{equation*}
\begin{cases}
(-q,\alpha)+k\ge 0, & s(\alpha)\in\Delta_+,\\
(-q,\alpha)+k>0, & s(\alpha)\in\Delta_-.
\end{cases}
\end{equation*}
This applies to both counting problems associated with $L_{u/m}$ and $S_{u/m}$.
Equivalently, the denominator on the set $S_{u/m}$ counts the number of elements in $S_{u/m}$ that are sent to positive affine roots:
\begin{equation*}
\tilde w(S_{u/m})\subset\tilde\Delta_+,
\end{equation*}

For $f=\text{regular}$ (the trivial parabolic $P$, so $\rho_P(\alpha)=0$ on positive roots), the relevant counts in formula~\ref{eq:hessenberg-dim} are
\begin{equation*}
\boxed{NumberA=\#\{(\alpha,k)\in L_{u/m}\mid (-q,\alpha)+k<0\}},
\end{equation*}
and
\begin{equation*}
\boxed{NumberB=\#\{(\alpha,k)\in S_{u/m}\mid (-q,\alpha)+k\le 0\}}.
\end{equation*}
These conditions do not depend on the finite Weyl-group part, so they are invariant under right multiplication by finite Weyl-group elements.
The dimension of the Hessenberg variety is given by $\mathrm{NumberA}-\mathrm{NumberB}$.

\subsection{Summary}\label{subsec:recap}
We summarize how to compute the fixed varieties (the associated affine Weyl group element $\tilde{w}=st_q$) for the input $((\mathfrak{g},o),\nu=\frac{u}{m},f)$, where $f$
gives rise to a standard parabolic subalgebra $P$ and an associated Weyl vector $\rho_P$:
\begin{enumerate}
\item Compute the sets $L_\nu$ and $S_\nu$, which consist of affine roots.
\item Count affine Weyl group elements $\tilde{w}$ that give rise to non-negative dimension, using formulas~\ref{equ1} and~\ref{equ2}.
  Check whether the corresponding cell $\tilde{w}^{-1}(\Delta_0)$ lies in a bounded region of the hyperplane arrangement determined by $S_\nu$.
 \item For the surviving set from the previous step, check the Chern-class constraints to ensure the cell is non-empty.
\item Group the solutions from step 2 by the coset $W_\nu\backslash\tilde W/W_P$, where $W_\nu$ is the group generated by reflections in $L_\nu$, and $W_P$ is
the group generated by the parabolic $P$. The group $W_\nu$ acts on the right, while the group $W_P$ acts on the left on the affine group element $\tilde{w}$.
\end{enumerate}
For more details on computing fixed varieties, see the Appendix C.

\section{\texorpdfstring{$W$-algebras}{W-algebras}}
Detailed constructions of $W^{k}(\mathfrak g,f)$ can be found in \cite{Frenkel:1992ju,deBoer:1993iz,kac2003quantum}, via generalized Drinfeld--Sokolov reduction of affine Kac--Moody algebras.
We summarize the ingredients used in what follows. The level $k$ of the $W$-algebra is $k=-h^\vee+\frac{1}{n}\frac{m}{u}$.
In our setup, $m=h^\vee$ gives boundary-admissible levels, while most examples of interest have $m\neq h^\vee$ and are non-admissible.

To each nilpotent orbit $f$ one associates an $\mathfrak{sl}_2$ triple $(e,f,x)$.
The semisimple element $x$ induces a half-integer grading
\begin{equation*}
\mathfrak g=\bigoplus_{l\in\frac{1}{2}\bbZ}\mathfrak g_l.
\end{equation*}
For $f=0$, this reduces to the affine Kac--Moody case.
The central charge is
\begin{equation}
\boxed{c\!\bigl(W^{k}(\mathfrak g,f)\bigr)=\dim \mathfrak g_0-\frac{1}{2}\dim \mathfrak g_{\frac12}-\frac{12}{k+h^\vee}\left|\rho-(k+h^\vee)x\right|^2},
\end{equation}
where $h^\vee$ is the dual Coxeter number.
The semisimple element $x$ has Dynkin labels $[a_1,\ldots,a_r]$ defined by $a_i=\alpha_i(x)$.
For a weight $\mu=\sum_i \mu_i\omega_i$, the norm is defined using the Cartan matrix:
\begin{equation*}
(\mu,\mu)=\sum_{i,j}F_{ij}\mu_i\mu_j,\qquad F_{ij}=(A^{-1})_{ij}\frac{\alpha_j^2}{2}.
\end{equation*}
The relevant data for two important choices of $f$ are listed in Table~\ref{tbl:f_cases}.
Recall
\begin{equation*}
\rho^2=\frac{h^\vee\,\dim\mathfrak g}{12}, \dim\mathfrak g=r(h+1)
\end{equation*}
\begin{table}[htbp]
    \centering
    \begin{tabular}{llll}
    \hline
    Case of \(f\) & Coordinate of \(x\) & \(\dim \mathfrak g_0\) & \(\dim \mathfrak g_{\frac{1}{2}}\) \\
    \hline
    \(f=0\) & \(0\) & \(\dim\mathfrak g\) & \(0\) \\
    \(f\) regular & \(\rho^\vee\) & \(r_G\) & \(0\) \\
    \hline
    \end{tabular}
    \caption{Coordinates of \(x\) and graded dimensions for two important cases of \(f\).}
    \label{tbl:f_cases}
    \end{table}

One also obtains modules by Drinfeld--Sokolov reduction of highest-weight representations of the affine Kac--Moody algebra with affine highest weight $\tilde\lambda$ (here we use the $-$ reduction functor \cite{Frenkel:1992ju,Arakawa2012Rationality}).
Their conformal weight is
\[
\boxed{
h(\lambda)
=
\frac{(\lambda\mid\lambda+2\rho)}{2(k+h^\vee)}
-\frac{k+h^\vee}{2}(x\mid x)
+(\rho\mid x).
}
\]
where $\lambda$ is the finite part of $\tilde\lambda$.
In particular,
\begin{align}\label{scale}
&f=0:\qquad c=\frac{k\,\dim\mathfrak g}{k+h^\vee},\qquad
h_\lambda=\frac{(\lambda\mid\lambda+2\rho)}{2(k+h^\vee)},\\
&f=\mathrm{regular}:\qquad
c=r-\frac{12}{um}(u\rho-m\rho^\vee)^2,\\[2pt]
&\hspace{4em}
c=-\frac{r}{um}\bigl((h+1)m-h^\vee u\bigr)
~~~\bigl(n\,h^\vee_{L^g}m-(h+1)u\bigr),\\[4pt]
&~~~~~~~~~~h_\lambda=\frac{(\lambda\mid\lambda+2\rho)}{2(k+h^\vee)}-\frac{m}{2u}(\rho^\vee)^2+(\rho,\rho^\vee).
\end{align}
In the regular case, we used the fact $k=-h^\vee+\frac{m}{u}$ and $h^\vee_{L^g}$
is the dual Coxeter number of the Langlands dual algebra. (For self-dual cases, such as $G_2$, use the usual dual Coxeter.)
$n$ is the ratio of the squared lengths of long and short roots,
which is 1 for simply-laced algebras and greater than 1 for non-simply-laced algebras. $\rho$ and $\rho^\vee$
are the Weyl and co-Weyl vectors of the Lie algebra $\mathfrak g$, respectively.

Given a simple $W$-algebra module $M$, its character is defined as
\begin{equation*}
I_M(q)=\tr_M q^{L_0-\frac{c}{24}}=q^{h-c/24}(a_0+\cdots).
\end{equation*}
where $h$ is the scaling dimension of the module $M$.
For the vacuum module, assuming a unique vacuum with $h=0$, the character has expansion
\begin{equation*}
I_0(q)=q^{-\frac{c}{24}}(1+aq+\cdots).
\end{equation*}
If the $W$-algebra contains an affine Kac--Moody subalgebra $J$, then each state of $M$ carries additional Cartan grading, yielding a refined character.

The $SL(2,\bbZ)$ action on characters is important for at least two reasons:
\begin{enumerate}
\item Modular invariance is essential for constructing modular-invariant partition functions of 2d CFTs.
\item Fusion rules can be extracted from the $S$ and $T$ matrices via the Verlinde formula.
\end{enumerate}

The following are some important classes of models for which modular properties of module characters are known.
For VOAs satisfying a finiteness condition, Zhu proved modular invariance of characters of irreducible modules \cite{zhu1996modular}.
For admissible affine Kac--Moody algebras, Kac--Wakimoto showed that admissible characters also satisfy modular transformation properties \cite{Kac:1988qc}.
For $C_2$-cofinite theories (reviewed below), characters of simple modules are not sufficient for modular closure; one must include logarithmic modules and their associated generalized characters \cite{miyamoto2004modular}.

\subsection{\texorpdfstring{$C_2$-cofinite $W$-algebra}{C2-cofinite W-algebra}}
It is generally difficult to determine all representations of a VOA directly from its intrinsic structure.
Zhu introduced an associative algebra $Zhu(V)$ and proved that many representation-theoretic questions for $V$ reduce to algebraic questions for $Zhu(V)$ \cite{zhu1996modular}.
In particular, when $Zhu(V)$ is finite-dimensional and semisimple, the VOA has strong finiteness properties, including finitely many simple modules and modular behavior of characters.
Such VOAs are called rational.

Zhu also defined the broader finiteness condition of $C_2$-cofiniteness.
For $C_2$-cofinite VOAs, the space of one-point functions still has modular properties and is spanned by characters of simple modules together with logarithmic modules \cite{miyamoto2004modular}.

In contrast to rational VOAs, $C_2$-cofinite VOAs typically have finite-dimensional but non-semisimple Zhu algebras, and their representation theory includes logarithmic modules \cite{miyamoto2004modular}.
As shown in \cite{miyamoto2004modular}, the space of one-point functions is generated as follows.

Let $V$ be a vertex operator algebra satisfying Zhu's finiteness condition, and let $c_V$ be the central charge of $V$.
Suppose every simple $V$-module is infinite-dimensional. Then the relevant vector space has a basis $s_{r,i_r}$, $0\le i_r\le k_r$, and
\begin{equation}
S^{r,i_r}(\tau)=\sum_{j=0}^{i_r}\sum_{k=0}^{\infty}S_{j,k}^{r,i_r}\,q^{r-\frac{c_V}{24}}(2\pi i\tau)^j.
\label{logchara}
\end{equation}
Here $r$ runs over conformal weights of simple modules, and $S_{0,0}^{r,i_r}\neq 0$ gives the non-logarithmic term.
Thus one can associate logarithmic modules of conformal weight $r$ through these generalized characters.

In general, computing Zhu's algebra and deciding $C_2$-cofiniteness are both difficult.
Arakawa introduced the associated variety attached to a VOA; a useful finiteness criterion is that this variety be zero-dimensional.
VOAs with this property are called lisse and are closely related to $C_2$-cofiniteness \cite{arakawa2012remark}.

Using the 4d/2d map, the associated variety of the VOA maps to the Higgs branch of the 4d theory.
One can then use independent 4d techniques to compute the associated variety, as in \cite{Xie:2019vzr}.
This gives a large class of lisse $W$-algebras whose representation theory we aim to study.
Guided by the 4d/2d correspondence, we use affine Springer geometry to formulate predictions for their VOA representation theory.

\section{Regular and subregular cases}

\subsection{Regular slope}

Consider the regular case, where the slope takes the form $\nu=\frac{u}{h^\vee}$ for the ADE case and $\nu=\frac{u}{h_\theta}$ for the twisted case.
The set $L_\nu$ is empty, and $S_\nu$ consists of the affine roots
\begin{equation*}
\{\alpha_1,\ldots, \alpha_r,\, -\theta+u\delta \},
\end{equation*}
where $\theta$ is the highest root. (The twisted case is treated separately in a later section.)
The fixed varieties are zero-dimensional and consist of a single point. The counting of affine Weyl group elements $\tilde{w}$ is then quite simple:
one only needs to find those $\tilde{w}$ for which the expected dimension of the fixed variety is zero; the Chern constraint is always satisfied.

When $f$ is regular, the relevant coset space is $\tilde{W}/W$, and it suffices to study orbit representatives of the form $t_\beta$.
The condition on $\beta$ is simply
\begin{equation*}
k-(\beta,\alpha)>0 \qquad\text{for all } (\alpha,k)\in S_\nu.
\end{equation*}
This set is the same as the one studied in the $W$-algebra context \cite{Frenkel:1992ju}.
The analogous condition for a general $f$ can be written down straightforwardly.

\subsection{Subregular case}

In the subregular case\footnote{This terminology comes from the fact that the associated irregular singularity is related to a subregular nilpotent orbit.} the slope again takes the form $\nu=\frac{u}{m}$.
The dual VOA is a $W$-algebra for the following values:
$m=4$ for $D_4$,
$m=9$ for $E_6$, $m=14$ for $E_7$, and $m=24$ for $E_8$.
Here $L_\nu$ is non-empty: it consists of two elements $\pm(\alpha,k)$, and the flag variety is $G_\nu/B$, whose coinvariant algebra is $\mathbb{C}[t_1]/t_1^2$,
where $t_1$ is the dual of the root $\alpha$.
The fixed varieties have dimension either one or zero.

We now consider the case where $f$ is regular, so the affine Weyl group element representing the double coset $W_{\nu}\backslash\tilde{W}/W$ is $t_\beta$ (up to the right action of $W_\nu$ generated by the affine roots in $L_\nu$).
\begin{enumerate}
\item $\dim=1$: this implies $\mathrm{NumberA}=1$, $\mathrm{NumberB}=0$. The Chern class of $V/F_t^yV$ is trivial, so the Chern constraint is automatically satisfied. The Hessenberg variety is $\mathbb{P}^1$ in this case.
\item $\dim=0$ with $\mathrm{NumberA}=1$, $\mathrm{NumberB}=1$: there is a single element $(\alpha_1,k)$ in $V/F_t^yV$, and its Chern class is $(\alpha_1,\alpha)\,t_1$. This vanishes if and only if $(\alpha_1,\alpha)=0$.
\item $\dim=0$ with $\mathrm{NumberA}=0$, $\mathrm{NumberB}=0$: the Chern constraint is automatically satisfied.
\end{enumerate}
The counting procedure is therefore as follows: first find all $\tilde{w}$ for which the Hessenberg variety has non-negative dimension; then, among those, it suffices to check the Chern constraint only for the case $\mathrm{NumberA}=\mathrm{NumberB}=1$ (one does not need to do this check if there is no such $\alpha_1$ in $S_\nu$ orthogonal to the root $\alpha$ in $L_\nu$).
See the Appendix for a nontrivial example ($\mathfrak{g}=E_6$, $\nu=11/9$, $f$~regular).

\newpage
\section{\texorpdfstring{Untwisted theories}{Untwisted theories}}

\subsection{\texorpdfstring{$D_4$}{D4}}

\subsubsection{\texorpdfstring{$m=4$}{m=4}}
We begin by reviewing the case already studied in \cite{shan2026mirror}.
Take $\mathfrak g=D_4$ with slope $\nu=\frac{1}{4}$. The corresponding 4d theory is $SU(2)$ gauge theory coupled to four hypermultiplets.
Its 4d central charge is $c_{4d}=\frac{7}{6}$, and its flavor symmetry is $SO(8)$.

To compute fixed varieties, we first determine $L_\nu$ and $S_\nu$.
In this case,
\begin{equation}
L_\nu=\left\{\alpha+l\delta\ \middle|\ \frac{1}{4}(\alpha,\rho^\vee)+l=0\right\}
\end{equation}
is non-empty:
\begin{equation}
L_\nu=\{\pm(-\mu+\delta) \},
\end{equation}
where $\mu=\alpha_1+\alpha_2+\alpha_3+\alpha_4$, so $W_\nu$ is the Weyl group generated by $s_{-\mu+\delta}$.
The set
\begin{equation}
S_\nu=\left\{\alpha+l\delta\ \middle|\ \frac{1}{4}(\alpha,\rho^\vee)+l=\frac{1}{4}\right\}
\end{equation}
is larger than in the admissible case:
\begin{equation}
S_\nu=\{\alpha_1,\alpha_2,\alpha_3,\alpha_4,-\mu+\alpha_1+\delta,-\mu+\alpha_3+\delta, -\mu+\alpha_4+\delta, \theta-\delta \},
\end{equation}
where $\theta=\mu+\alpha_2=\alpha_1+2\alpha_2+\alpha_3+\alpha_4$ is the highest root. We adopt the Bourbaki numbering for simple roots \cite{bourbaki2006groupes,carter2005lie}.

The affine Weyl-group elements take the form $\tilde{w}=st_q$.
The $\bbZ_2$ subgroup generated by $s_{-\mu+\delta}$ acts on $t_q$.
In this example, there is a single one-dimensional fixed variety, while all remaining fixed loci are isolated points; see Table~\ref{tab:d4_lnu_pairs} for the list of solutions.
\begin{table}[htbp]
\centering
\scriptsize
\setlength{\tabcolsep}{4pt}
\begin{tabular}{c|c|c|c}
\hline
Pair &
Left solution $(s;\,m_1,m_2,m_3,m_4)$ &
Right solution $(s';\,m_1',m_2',m_3',m_4')$ &
$\mathrm{NumberA}-\mathrm{NumberB}$ \\
\hline
1 &
$(e;\,0,0,0,0)$ &
$(s_1s_3s_2s_4s_2s_1s_3;\,-1,-1,-1,-1)$ &
$0$ \\
2 &
$(s_2s_1s_3s_2s_4s_2s_1;\,-1,-2,-1,-1)$ &
$(s_2s_1s_3s_2s_1s_4s_2s_1s_3s_2;\,-1,-2,-1,-1)$ &
$0$ \\
3 &
$(s_2s_1s_3s_2s_4s_2s_3;\,-1,-2,-1,-1)$ &
$(s_1s_2s_1s_3s_2s_4s_2s_1s_3s_2;\,-1,-2,-1,-1)$ &
$0$ \\
4 &
$(s_2s_1s_3s_4s_2s_1s_3;\,-1,-2,-1,-1)$ &
$(s_2s_1s_3s_2s_4s_2s_1s_3s_2s_4;\,-1,-2,-1,-1)$ &
$0$ \\
5 &
$(s_2s_1s_3s_2s_4s_2s_1s_3;\,-1,-2,-1,-1)$ &
$(s_2s_1s_3s_2s_4s_2s_1s_3s_2;\,-1,-2,-1,-1)$ &
$1$ \\
\hline
\end{tabular}
\caption{Pairs of $D_4$ solutions related by the right action of the affine reflection generated by $L_\nu=\{\pm(-\mu,1)\}$.}
\label{tab:d4_lnu_pairs}
\end{table}

\begin{table}[htbp]
    \centering
    \scriptsize
    \begin{tabular}{c|l|l}
    \hline
    \# & $(s;\,m_1,m_2,m_3,m_4)$ & $\tilde w(-2\Lambda_0+\tilde\rho)-\tilde\rho$ \\
    \hline
    1 & $(e;\,0,0,0,0)$ & $-2\Lambda_0$ \\
    2 & $(s_1s_3s_2s_4s_2s_1s_3;\,-1,-1,-1,-1)$ & $-2\Lambda_0$ \\
    3 & $(s_2s_1s_3s_2s_4s_2s_1;\,-1,-2,-1,-1)$ & $\delta-2\Lambda_3$ \\
    4 & $(s_2s_1s_3s_2s_4s_2s_3;\,-1,-2,-1,-1)$ & $\delta-2\Lambda_1$ \\
    5 & $(s_2s_1s_3s_4s_2s_1s_3;\,-1,-2,-1,-1)$ & $\delta-2\Lambda_4$ \\
    6 & $(s_2s_1s_3s_2s_4s_2s_1s_3;\,-1,-2,-1,-1)$ & $\delta-\Lambda_2$ \\
    7 & $(s_2s_1s_3s_2s_4s_2s_1s_3s_2;\,-1,-2,-1,-1)$ & $\delta-\Lambda_2$ \\
    8 & $(s_1s_2s_1s_3s_2s_4s_2s_1s_3s_2;\,-1,-2,-1,-1)$ & $\delta-2\Lambda_1$ \\
    9 & $(s_2s_1s_3s_2s_1s_4s_2s_1s_3s_2;\,-1,-2,-1,-1)$ & $\delta-2\Lambda_3$ \\
    10 & $(s_2s_1s_3s_2s_4s_2s_1s_3s_2s_4;\,-1,-2,-1,-1)$ & $\delta-2\Lambda_4$ \\
    \hline
    \end{tabular}
    \caption{Values of $\tilde w(-2\Lambda_0+\tilde\rho)-\tilde\rho$ for the 10 solutions associated with the fixed varieties of the affine Springer fiber.}
    \end{table}

We now review the characters of the simple representations of $D_{-2}(D_4)$; see \cite{zheng2022surface}.
The simple characters computed in \cite{zheng2022surface} are not closed under the modular $S$ transformation.
To obtain modular closure, one must include logarithmic characters.
This is naturally explained by the modular differential equation.

The modular differential equation for characters of the $D_4$ theory \cite{arakawa2018quasi,kaneko2013modular} is
\begin{equation*}
f''(\tau)-\frac{k+1}{6}E_2(\tau)f'(\tau)+\frac{k(k+1)}{12}E_2'(\tau)f(\tau)=0
\end{equation*}
where $E_2$ is the Eisenstein series and the derivative is defined by
\[
'=q\frac{d}{dq}=\frac{1}{2\pi i}\frac{d}{d\tau}.
\]
For our theory, $k=5$. There is a vacuum solution of the form $I(q)=q^{\alpha}(1+\ldots)$ with $\alpha=\frac{k+2}{12}$.
This solution is quasi-modular rather than modular.
There is also a logarithmic solution
\begin{equation*}
    g_1\log q+ g_2
\end{equation*}
with expansion
\begin{equation*}
g_1=q^{\alpha}(1+\ldots),\qquad g_2=q^\beta(b_0+\ldots)
\end{equation*}
In the current case $\alpha=\frac{7}{12}, \beta=-\frac{5}{12}$. Comparing with the character formula for logarithmic modules (see \eqref{logchara}), this module should be associated with a simple module with scaling dimension $\beta=r-\frac{c}{24}\to r=-1$. This scaling dimension is indeed the one
corresponding to the one-dimensional fixed variety.

We now summarize how to interpret VOA modules from the geometry of the affine Springer fiber:
\begin{enumerate}
\item Each non-empty cell gives rise to a simple module whose highest weights are computed by \eqref{eq:highest-weight-map} (there may be more than one simple module depending on the irreducible components associated with a fixed variety).
\item Because higher-dimensional fixed varieties appear, the affine Springer cohomology exceeds the naive cell count.
In the present case, a one-dimensional fixed variety contributes two dimensions to cohomology.
Hence the total cohomology dimension is $6$, matching the number of characters required on the VOA side.
\end{enumerate}
This leads to the following proposal for predicting modules of the $W$-algebra from affine Springer cohomology:
\begin{enumerate}
\item The dimension of affine Springer cohomology \footnote{The precise statement should be the part corresponding to the irreducible representation of DAHA.} matches the number of VOA characters needed for modular closure.
\item Each higher-dimensional fixed variety labeled by $\tilde{w}$ contributes simple modules and additional logarithmic modules whose number is given by the dimension of the cohomology associated with the fixed varieties.
\end{enumerate}
Recall that affine Springer cohomology carries a DAHA representation, which automatically has modular properties \cite{cherednik2005double}.
The above identification therefore explains the modular behavior of the corresponding VOA.

Consider the general case with data $\mathfrak g=D_4,\;\nu=\frac{u}{4}$. The sets $S_\nu$ and $L_\nu$ are given as
\begin{equation}
    S_\nu=\{\alpha_1,\alpha_2,\alpha_3,\alpha_4,-\mu+\alpha_1+u\delta,-\mu+\alpha_3+u\delta,-\mu+\alpha_4+u\delta,\theta-u\delta\},
    \end{equation}
    \begin{equation}
        L_\nu=\{\pm(-\mu+u\delta)\} .
        \end{equation}

In this case, we first compute affine Weyl group elements $\tilde{w}$ that have non-negative expected dimension. The dimension formula
can be written as $\mathrm{NumberA}-\mathrm{NumberB}$, where $\mathrm{NumberA}$ involves the space $L_\nu$ and $\mathrm{NumberB}$ involves the space $S_\nu$.
Since $\mathrm{NumberA}$ is always one, $\mathrm{NumberB}$ must be at most one to obtain a non-negative dimension. The condition $\mathrm{NumberB}\le 1$ means that
at most one element has negative sign for the region $\tilde{w}^{-1}(\Delta_0)$. When all signs are positive, the region is always bounded; one computes the recession cone
defined as the solution space $\{v\mid \operatorname{sgn}_i\beta_i(v)\ge 0\}$, where $\beta_i$ are the finite parts of the affine roots in $S_\nu$. Looking at $S_\nu$, there are four negative finite roots and four positive finite roots; for any choice of signs of these roots, the recession cone
has only the zero solution. We conclude that for any affine Weyl group element $\tilde{w}$ giving rise to non-negative expected dimension,
the corresponding cell is non-empty. The Chern constraint is also automatically satisfied, which makes the counting problem much simpler.

For $f=0$, one can use the boundedness criterion~\eqref{eq:cell-nonempty} to compute the fixed varieties.
The number of affine Weyl group elements that give rise to non-empty cells is $10u^4$; modulo the $\bbZ_2$ action, this becomes $5u^4$ solutions.
This suggests $5u^4$ irreducible modules in category $\mathcal O$ for the corresponding VOA (for $u=1$ see \cite{Arakawa_2016}, for $u=3$ see \cite{adamovic2025new}).

For $u=3$, there are 81 fixed varieties of dimension 1, and 324 fixed points, giving a total cohomology dimension of $486$. The highest weights
on the VOA side are computed by the formula $\tilde{w}(-14/3\Lambda_0+\tilde{\rho})-\tilde{\rho}$, and this matches the result found in \cite{adamovic2025new}
(our computed set of weights matches the results in \cite{adamovic2025new}). On the VOA side, we have the affine Lie algebra $V_{-14/3}(D_4)$,
and based on the computation from the affine Springer fiber side, one predicts that 486 characters are needed for modular closure, including 405 simple characters and 81 logarithmic characters.
Our approach is quite powerful: we can compute the number of irreducible modules for any value of $u$ straightforwardly, while on the VOA side one needs to compute complicated singular vectors.

Next, consider $f=\mathrm{regular}$, for which the corresponding $W$-algebra should be $C_2$-cofinite. The sets $L_\nu$ and $S_\nu$ do not depend on $f$, and thus still take the same form as in the case $f=0$.

For $u=5$ and $f=\mathrm{regular}$, the physical theory has just one Coulomb branch operator with scaling dimension $\frac{6}{5}$.
This is expected to be the same theory as the $(A_1,A_2)$ theory, and the corresponding VOA is the $(2,5)$ Virasoro minimal model, which is rational and
has just two simple modules. In the present case,
the corresponding VOA is the principal $W$-algebra
\begin{equation*}
V=W^{-\frac{26}{5}}(D_4,f_{\mathrm{prin}}) .
\end{equation*}
On the affine Springer fiber side, the fixed varieties can be computed using the dimension formula. The answer is labeled by the elements in the coset $W_\nu\backslash\tilde W/W$, so one only needs to determine the translation part $t_q$ of the coroot vectors. These are also related by the action of reflections generated by the affine roots in $L_\nu$. The final result is that there are only two fixed varieties, and both are fixed points (see Table~\ref{tab:d4_u5_orbit_weight_dim}). This matches the result for the $(2,5)$ Virasoro minimal model. The scaling dimensions of the two modules, computed using the map~\eqref{eq:highest-weight-map} and formula~\ref{scale}, as well as the central charge of this model, also agree with the Virasoro $(2,5)$ minimal model.
\begin{table}[htbp]
    \centering
    \small
    \setlength{\tabcolsep}{4pt}
    \renewcommand{\arraystretch}{1.15}
    \begin{tabular}{c|c|c|c|c}
    \hline
    Orbit & $(m_1,m_2,m_3,m_4)$ & $(\omega_1,\omega_2,\omega_3,\omega_4)$ & $\mathrm{dim}$ & $h_{\lambda}$ \\
    \hline
    $\mathcal O_1$
    & $(-4,-7,-4,-4)$
    & $(-\frac{4}{5},-\frac{8}{5},-\frac{4}{5},-\frac{4}{5})$
    & $0$
    & $-\frac{1}{5}$ \\
    \hline
    $\mathcal O_2$
    & $(-4,-6,-4,-4)$
    & $(-\frac{8}{5},0,-\frac{8}{5},-\frac{8}{5})$
    & $0$
    & $0$ \\
    \hline
    \end{tabular}
    \caption{Orbit decomposition for slope $\nu=5/4$.  Listed are the finite $D_4$ weights $(m_i)$ (which are coefficients in root basis) and $(\omega_i)$ (the coefficients in weight basis), the dimension
    $\mathrm{dim}=\mathrm{NumberA}-\mathrm{NumberB}$, and the conformal weight $h_{\lambda}$.}
    \label{tab:d4_u5_orbit_weight_dim}
    \end{table}

For $u=7$ and $f=\mathrm{regular}$, the $W$-algebra is
\begin{equation*}
    W^{-\frac{38}{7}}(D_4,f_{\mathrm{prin}})
\end{equation*}
\par
Our method yields the following fixed varieties:
one one-dimensional variety and
16 zero-dimensional varieties.
The prediction is that there are 17 simple modules and one logarithmic module for this algebra; see Table~\ref{tab:d4_u7_orbit_weight_dim}.

For $u=9$, there are $66$ zero-dimensional fixed varieties and $6$ one-dimensional fixed varieties. For $u=11$, there are $160$ zero-dimensional fixed points
and $20$ one-dimensional varieties. Based on our computations, we conjecture that the total dimension of the cohomology groups is given by the formula
\begin{equation*}
    \frac{1}{32}(u-3)^2(u-1)^2 .
\end{equation*}
The numbers of one-dimensional and zero-dimensional fixed varieties are given by
\[
N_1(u)=\frac{(u-1)(u-3)^2(u-5)}{192},\qquad
N_0(u)=\frac{(u-1)(u-3)^2(u+1)}{48},
\]

    \begingroup
    \scriptsize
    \setlength{\tabcolsep}{6pt}
    \renewcommand{\arraystretch}{1.1}
    \begin{longtable}{c|c|c|c}
    \caption{Orbit decomposition for $D_4$, slope $\nu=7/4$.  Listed are the coroot vectors $\beta$, the expected dimension
    $\mathrm{dim}=\mathrm{NumberA}-\mathrm{NumberB}$, and the conformal weight $h_{\lambda}$.}
    \label{tab:d4_u7_orbit_weight_dim}\\
    \hline
    Orbit & $\beta$ & $\mathrm{dim}$ & $h_{\lambda}$ \\
    \hline
    \endfirsthead
    \hline
    Orbit & $\beta$ & $\mathrm{dim}$ & $h_{\lambda}$ \\
    \hline
    \endhead

    $\mathcal O_1$ & $(-7,-11,-7,-6)$ & $0$ & $-1$ \\
     & $(-5,-9,-5,-4)$ & $0$ & $-1$ \\
    \hline

    $\mathcal O_2$ & $(-7,-11,-6,-7)$ & $0$ & $-1$ \\
     & $(-5,-9,-4,-5)$ & $0$ & $-1$ \\
    \hline

    $\mathcal O_3$ & $(-6,-11,-7,-7)$ & $0$ & $-1$ \\
     & $(-4,-9,-5,-5)$ & $0$ & $-1$ \\
    \hline

    $\mathcal O_4$ & $(-6,-11,-6,-6)$ & $0$ & $-\frac{9}{7}$ \\
    \hline

    $\mathcal O_5$ & $(-6,-10,-6,-6)$ & $1$ & $-2$ \\
     & $(-5,-9,-5,-5)$ & $1$ & $-2$ \\
    \hline

    $\mathcal O_6$ & $(-6,-9,-6,-6)$ & $0$ & $-\frac{11}{7}$ \\
     & $(-4,-7,-4,-4)$ & $0$ & $-\frac{11}{7}$ \\
    \hline

    $\mathcal O_7$ & $(-6,-9,-6,-5)$ & $0$ & $-\frac{12}{7}$ \\
     & $(-5,-8,-5,-4)$ & $0$ & $-\frac{12}{7}$ \\
    \hline

    $\mathcal O_8$ & $(-6,-9,-5,-6)$ & $0$ & $-\frac{12}{7}$ \\
     & $(-5,-8,-4,-5)$ & $0$ & $-\frac{12}{7}$ \\
    \hline

    $\mathcal O_9$ & $(-6,-9,-5,-5)$ & $0$ & $-\frac{13}{7}$ \\
    \hline

    $\mathcal O_{10}$ & $(-6,-8,-6,-6)$ & $0$ & $0$ \\
     & $(-3,-5,-3,-3)$ & $0$ & $0$ \\
    \hline

    $\mathcal O_{11}$ & $(-6,-8,-5,-5)$ & $0$ & $-\frac{10}{7}$ \\
     & $(-5,-7,-4,-4)$ & $0$ & $-\frac{10}{7}$ \\
    \hline

    $\mathcal O_{12}$ & $(-5,-9,-6,-6)$ & $0$ & $-\frac{12}{7}$ \\
     & $(-4,-8,-5,-5)$ & $0$ & $-\frac{12}{7}$ \\
    \hline

    $\mathcal O_{13}$ & $(-5,-9,-6,-5)$ & $0$ & $-\frac{13}{7}$ \\
    \hline

    $\mathcal O_{14}$ & $(-5,-9,-5,-6)$ & $0$ & $-\frac{13}{7}$ \\
    \hline

    $\mathcal O_{15}$ & $(-5,-8,-6,-5)$ & $0$ & $-\frac{10}{7}$ \\
     & $(-4,-7,-5,-4)$ & $0$ & $-\frac{10}{7}$ \\
    \hline

    $\mathcal O_{16}$ & $(-5,-8,-5,-6)$ & $0$ & $-\frac{10}{7}$ \\
     & $(-4,-7,-4,-5)$ & $0$ & $-\frac{10}{7}$ \\
    \hline

    $\mathcal O_{17}$ & $(-5,-8,-5,-5)$ & $0$ & $-\frac{15}{7}$ \\
    \hline

    \end{longtable}
    \endgroup

\subsubsection{\texorpdfstring{$m=2$}{m=2}}
We consider the slope $\nu=\frac{u}{2}$. The sets $L_\nu$ and $S_\nu$ now consist of the following affine roots $(\alpha,k)$:
    \[
    \begin{aligned}
    L_{u/2}
    =\{&
    (\alpha_1+\alpha_2,\,-u),\;
    (\alpha_2+\alpha_3,\,-u),\;
    (\alpha_2+\alpha_4,\,-u),\;
    (\alpha_1+\alpha_2+\alpha_3+\alpha_4,\,-2u),\\
    &(-\alpha_1-\alpha_2,\,u),\;
    (-\alpha_2-\alpha_3,\,u),\;
    (-\alpha_2-\alpha_4,\,u),\;
    (-\alpha_1-\alpha_2-\alpha_3-\alpha_4,\,2u)
    \}.
    \end{aligned}
    \]
    \[
    \begin{aligned}
    S_{u/2}
    =\{&
    (\alpha_1,0),\;(\alpha_2,0),\;(\alpha_3,0),\;(\alpha_4,0),\\
    &(\alpha_2+\alpha_3+\alpha_4,\,-u),\;
    (\alpha_1+\alpha_2+\alpha_3,\,-u),\;
    (\alpha_1+\alpha_2+\alpha_4,\,-u),\\
    &(\alpha_1+2\alpha_2+\alpha_3+\alpha_4,\,-2u),\\
    &(-\alpha_1,\,u),\;(-\alpha_2,\,u),\;(-\alpha_3,\,u),\;(-\alpha_4,\,u),\\
    &(-\alpha_2-\alpha_3-\alpha_4,\,2u),\;
    (-\alpha_1-\alpha_2-\alpha_3,\,2u),\;
    (-\alpha_1-\alpha_2-\alpha_4,\,2u),\\
    &(-\alpha_1-2\alpha_2-\alpha_3-\alpha_4,\,3u)
    \}.
    \end{aligned}
    \]
    We can then compute the fixed varieties for $f=\mathrm{regular}$. For $u=3$, there is one zero-dimensional variety and two one-dimensional varieties; see Table~\ref{tab:d4-m2-u3}. This theory is described by the singularity $x^2+y^3+z^3+w^3$,
which has a gauge-theory description as an $SU(2)$ gauge group coupled to three $D_3(SU(2))$ theories, with Coulomb branch spectrum $(2,\frac{4}{3},\frac{4}{3},\frac{4}{3})$. It is interesting to compare our prediction with the result in \cite{jiang2024modularity}.
For $u=5$, the counting is given in Table~\ref{tab:d4-m2-u5} and the pattern of numbers of varieties is as follows:

    \begin{longtable}{c|c|c|c}
        \caption{Orbit decomposition for $D_4$, slope $\nu=\frac{3}{2}$. Listed are the orbit representatives $\beta$, the expected dimension $\mathrm{dim}=\mathrm{NumberA}-\mathrm{NumberB}$, and the conformal weight $h_{\lambda}$.}
        \label{tab:d4-m2-u3}\\
        \hline
        Orbit & $\beta$ & $\mathrm{dim}$ & $h_{\lambda}$ \\
        \hline
        \endfirsthead
        \hline
        Orbit & $\beta$ & $\mathrm{dim}$ & $h_{\lambda}$ \\
        \hline
        \endhead
        $\mathcal O_{1}$ & $(-4,-6,-3,-3)$ & $0$ & $0$ \\
        $\mathcal O_{2}$ & $(-4,-6,-4,-4)$ & $1$ & $-\frac{2}{3}$ \\
        $\mathcal O_{3}$ & $(-5,-8,-5,-5)$ & $1$ & $-1$ \\
        \hline
        \end{longtable}

    \begin{longtable}{c|c|c|c}
        \caption{Orbit decomposition for $D_4$, slope $\nu=\frac{5}{2}$. Listed are the orbit representatives $\beta$, the expected dimension $\mathrm{dim}=\mathrm{NumberA}-\mathrm{NumberB}$, and the conformal weight $h_{\lambda}$. Orbits marked with $^{*}$ are eliminated by the Chern-class constraint (their top Chern product lies in the ideal $I_\beta$, so the Hessenberg variety is empty).}
        \label{tab:d4-m2-u5}\\
        \hline
        Orbit & $\beta$ & $\mathrm{dim}$ & $h_{\lambda}$ \\
        \hline
        \endfirsthead
        \hline
        Orbit & $\beta$ & $\mathrm{dim}$ & $h_{\lambda}$ \\
        \hline
        \endhead
        $\mathcal O_{1}$ & $(-6,-14,-6,-6)$ & $0$ & $0$ \\
        $\mathcal O_{2}$ & $(-7,-13,-5,-7)$ & $0$ & $-\frac{14}{5}$ \\
        $\mathcal O_{3}$ & $(-7,-14,-7,-6)$ & $0$ & $-\frac{14}{5}$ \\
        $\mathcal O_{4}$ & $(-9,-12,-9,-9)$ & $0$ & $-\frac{13}{5}$ \\
        $\mathcal O_{5}$ & $(-8,-15,-7,-7)$ & $0$ & $-3$ \\
        $\mathcal O_{6}$ & $(-4,-12,-8,-8)$ & $0$ & $-\frac{8}{5}$ \\
        $\mathcal O_{7}$ & $(-5,-7,-4,-4)$ & $0$ & $-\frac{14}{5}$ \\
        $\mathcal O_{8}$ & $(-6,-11,-6,-5)$ & $0$ & $-\frac{22}{5}$ \\
        $\mathcal O_{9}$ & $(-7,-12,-7,-5)$ & $0$ & $-\frac{21}{5}$ \\
        $\mathcal O_{10}$ & $(-10,-15,-8,-9)$ & $1$ & $-\frac{24}{5}$ \\
        $\mathcal O_{11}$ & $(-5,-12,-8,-8)$ & $0$ & $-\frac{19}{5}$ \\
        $\mathcal O_{12}$ & $(-6,-13,-6,-7)$ & $0$ & $-\frac{18}{5}$ $^{*}$ \\
        $\mathcal O_{13}$ & $(-5,-7,-5,-4)$ & $0$ & $-3$ \\
        $\mathcal O_{14}$ & $(-7,-11,-10,-6)$ & $0$ & $-\frac{8}{5}$ \\
        $\mathcal O_{15}$ & $(-8,-14,-9,-10)$ & $1$ & $-\frac{23}{5}$ \\
        $\mathcal O_{16}$ & $(-5,-11,-6,-8)$ & $1$ & $-4$ \\
        $\mathcal O_{17}$ & $(-10,-12,-7,-6)$ & $0$ & $-\frac{13}{5}$ $^{*}$ \\
        $\mathcal O_{18}$ & $(-6,-10,-4,-6)$ & $0$ & $-\frac{18}{5}$ $^{*}$ \\
        $\mathcal O_{19}$ & $(-5,-9,-5,-6)$ & $1$ & $-\frac{23}{5}$ \\
        $\mathcal O_{20}$ & $(-5,-13,-8,-7)$ & $0$ & $-3$ \\
        $\mathcal O_{21}$ & $(-8,-12,-5,-7)$ & $1$ & $-4$ \\
        $\mathcal O_{22}$ & $(-5,-11,-9,-5)$ & $0$ & $-\frac{8}{5}$ \\
        $\mathcal O_{23}$ & $(-6,-14,-10,-9)$ & $0$ & $-\frac{13}{5}$ $^{*}$ \\
        $\mathcal O_{24}$ & $(-7,-14,-8,-7)$ & $1$ & $-\frac{24}{5}$ \\
        $\mathcal O_{25}$ & $(-8,-12,-5,-8)$ & $0$ & $-\frac{19}{5}$ \\
        $\mathcal O_{26}$ & $(-8,-12,-8,-8)$ & $3$ & $-\frac{28}{5}$ \\
        $\mathcal O_{27}$ & $(-9,-15,-9,-8)$ & $1$ & $-\frac{27}{5}$ \\
        $\mathcal O_{28}$ & $(-7,-13,-6,-8)$ & $1$ & $-\frac{24}{5}$ \\
        $\mathcal O_{29}$ & $(-7,-11,-8,-7)$ & $1$ & $-\frac{27}{5}$ \\
        $\mathcal O_{30}$ & $(-9,-13,-7,-7)$ & $1$ & $-\frac{26}{5}$ \\
        $\mathcal O_{31}$ & $(-8,-14,-6,-9)$ & $0$ & $-\frac{19}{5}$ \\
        $\mathcal O_{32}$ & $(-9,-14,-10,-8)$ & $1$ & $-\frac{23}{5}$ \\
        $\mathcal O_{33}$ & $(-6,-8,-5,-5)$ & $1$ & $-4$ \\
        $\mathcal O_{34}$ & $(-9,-14,-6,-10)$ & $0$ & $-\frac{13}{5}$ $^{*}$ \\
        $\mathcal O_{35}$ & $(-6,-11,-6,-6)$ & $1$ & $-\frac{27}{5}$ \\
        $\mathcal O_{36}$ & $(-6,-13,-8,-8)$ & $2$ & $-5$ \\
        $\mathcal O_{37}$ & $(-6,-11,-9,-6)$ & $1$ & $-\frac{18}{5}$ \\
        $\mathcal O_{38}$ & $(-8,-13,-7,-7)$ & $2$ & $-\frac{29}{5}$ \\
        $\mathcal O_{39}$ & $(-8,-13,-9,-7)$ & $1$ & $-\frac{27}{5}$ \\
        $\mathcal O_{40}$ & $(-8,-15,-9,-8)$ & $1$ & $-\frac{26}{5}$ \\
        $\mathcal O_{41}$ & $(-8,-15,-8,-9)$ & $1$ & $-\frac{26}{5}$ \\
        $\mathcal O_{42}$ & $(-7,-12,-6,-5)$ & $0$ & $-\frac{18}{5}$ $^{*}$ \\
        $\mathcal O_{43}$ & $(-8,-12,-7,-8)$ & $2$ & $-\frac{29}{5}$ \\
        $\mathcal O_{44}$ & $(-8,-14,-8,-9)$ & $2$ & $-\frac{29}{5}$ \\
        $\mathcal O_{45}$ & $(-7,-13,-8,-8)$ & $3$ & $-6$ \\
        $\mathcal O_{46}$ & $(-7,-12,-7,-7)$ & $1$ & $-\frac{31}{5}$ \\
        \hline
        \end{longtable}

The pattern of numbers of non-empty varieties (after imposing the Chern-class constraint) in each dimension is as follows:
\[
\begin{aligned}
N_4(u) &= \frac{(u-1)(u-3)^2(u-5)}{192},\\[2pt]
N_3(u) &= \frac{(u-1)(u-3)^2(u+19)}{192},\\[2pt]
N_2(u) &= \frac{(u-1)(u-3)(u^2-4u+19)}{48},\\[2pt]
N_1(u) &= \frac{(u-1)(3u^3+7u^2-35u+57)}{96},\\[2pt]
N_0(u) &= \frac{(u-1)^4}{16}.
\end{aligned}
\]
The Chern-class constraint eliminates $\frac{(u-1)(u-2)(u-3)}{4}$ zero-dimensional orbits (those with $\mathrm{NumberA}=\mathrm{NumberB}=4$ whose four linear forms are linearly dependent, causing the top Chern product to lie in the ideal $I_\beta$).  The remaining dimensions are unaffected by the Chern test.  For $u=5$, six orbits are eliminated (marked with $^{*}$ in Table~\ref{tab:d4-m2-u5}), reducing the total from $46$ to $40$.  For $u=7$, $30$ orbits are eliminated, giving $189$ non-empty orbits; for $u=9$, $84$ are eliminated, giving $576$.

\subsection{\texorpdfstring{$E_6$}{E6}}

We consider the Lie algebra $\mathfrak{g}=E_6$ with slope $\nu=\frac{4}{3}$.
The dual Coxeter number is $h^\vee=12$, and the level is
\begin{equation*}
k=-h^\vee+\frac{m}{u}=-12+\frac{3}{4}=-\frac{45}{4}.
\end{equation*}

The sets $L_\nu$ and $S_\nu$ are determined by the conditions
\begin{equation*}
L_{4/3}:\quad 4(\alpha,\rho)+3k=0,
\qquad
S_{4/3}:\quad 4(\alpha,\rho)+3k=4.
\end{equation*}
Explicitly, $L_\nu$ consists of the following $18$ affine roots $(\alpha,k)$:
\[
\begin{aligned}
L_{4/3}=\{&
(\alpha_1+\alpha_3+\alpha_4,-4),\;
(\alpha_2+\alpha_3+\alpha_4,-4),\;
(\alpha_2+\alpha_4+\alpha_5,-4),\\
&(\alpha_3+\alpha_4+\alpha_5,-4),\;
(\alpha_4+\alpha_5+\alpha_6,-4),\\
&(\alpha_1+\alpha_2+\alpha_3+2\alpha_4+\alpha_5,-8),\;
(\alpha_1+\alpha_2+\alpha_3+\alpha_4+\alpha_5+\alpha_6,-8),\\
&(\alpha_2+\alpha_3+2\alpha_4+\alpha_5+\alpha_6,-8),\;
(\alpha_1+\alpha_2+2\alpha_3+2\alpha_4+2\alpha_5+\alpha_6,-12),\\
&(-\alpha_1-\alpha_3-\alpha_4,4),\;
(-\alpha_2-\alpha_3-\alpha_4,4),\;
(-\alpha_2-\alpha_4-\alpha_5,4),\\
&(-\alpha_3-\alpha_4-\alpha_5,4),\;
(-\alpha_4-\alpha_5-\alpha_6,4),\\
&(-\alpha_1-\alpha_2-\alpha_3-2\alpha_4-\alpha_5,8),\;
(-\alpha_1-\alpha_2-\alpha_3-\alpha_4-\alpha_5-\alpha_6,8),\\
&(-\alpha_2-\alpha_3-2\alpha_4-\alpha_5-\alpha_6,8),\;
(-\alpha_1-\alpha_2-2\alpha_3-2\alpha_4-2\alpha_5-\alpha_6,12)
\}.
\end{aligned}
\]
The set $S_\nu$ contains $27$ affine roots:
\[
\begin{aligned}
S_{4/3}=\{&
\alpha_1,\;\alpha_2,\;\alpha_3,\;\alpha_4,\;\alpha_5,\;\alpha_6,\\
&\alpha_1+\alpha_2+\alpha_3+\alpha_4-4\delta,\;
\alpha_1+\alpha_3+\alpha_4+\alpha_5-4\delta,\;
\alpha_2+\alpha_3+\alpha_4+\alpha_5-4\delta,\\
&\alpha_2+\alpha_4+\alpha_5+\alpha_6-4\delta,\;
\alpha_3+\alpha_4+\alpha_5+\alpha_6-4\delta,\\
&\alpha_1+\alpha_2+2\alpha_3+2\alpha_4+\alpha_5-8\delta,\;
\alpha_1+\alpha_2+\alpha_3+2\alpha_4+\alpha_5+\alpha_6-8\delta,\\
&\alpha_2+\alpha_3+2\alpha_4+2\alpha_5+\alpha_6-8\delta,\\
&\alpha_1+\alpha_2+2\alpha_3+3\alpha_4+2\alpha_5+\alpha_6-12\delta,\\
&-\alpha_1-\alpha_3+4\delta,\;
-\alpha_2-\alpha_4+4\delta,\;
-\alpha_3-\alpha_4+4\delta,\;
-\alpha_4-\alpha_5+4\delta,\;
-\alpha_5-\alpha_6+4\delta,\\
&-\alpha_1-\alpha_2-\alpha_3-\alpha_4-\alpha_5+8\delta,\;
-\alpha_1-\alpha_3-\alpha_4-\alpha_5-\alpha_6+8\delta,\\
&-\alpha_2-\alpha_3-2\alpha_4-\alpha_5+8\delta,\;
-\alpha_2-\alpha_3-\alpha_4-\alpha_5-\alpha_6+8\delta,\\
&-\alpha_1-\alpha_2-2\alpha_3-2\alpha_4-\alpha_5-\alpha_6+12\delta,\\
&-\alpha_1-\alpha_2-\alpha_3-2\alpha_4-2\alpha_5-\alpha_6+12\delta,\\
&-\alpha_1-2\alpha_2-2\alpha_3-3\alpha_4-2\alpha_5-\alpha_6+16\delta
\}.
\end{aligned}
\]

We adopt the Bourbaki numbering for the simple roots of $E_6$:
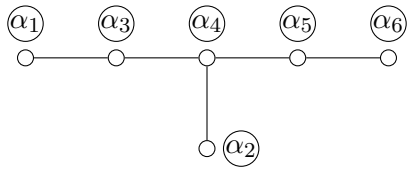
\begin{center}
\begin{tikzpicture}[scale=0.8]
\tikzset{every node/.style={circle,draw,minimum size=6pt,inner sep=0pt}}
\node (a1) at (0,0) {};
\node[above=6pt] at (a1) {$\alpha_1$};
\node (a3) at (1.5,0) {};
\node[above=6pt] at (a3) {$\alpha_3$};
\node (a4) at (3,0) {};
\node[above=6pt] at (a4) {$\alpha_4$};
\node (a5) at (4.5,0) {};
\node[above=6pt] at (a5) {$\alpha_5$};
\node (a6) at (6,0) {};
\node[above=6pt] at (a6) {$\alpha_6$};
\node (a2) at (3,-1.5) {};
\node[right=6pt] at (a2) {$\alpha_2$};
\draw (a1)--(a3)--(a4)--(a5)--(a6);
\draw (a4)--(a2);
\end{tikzpicture}
\end{center}

The finite Weyl group $W_\nu$ is generated by reflections in the $18$ roots of $L_\nu$.
A brute-force search over the coroot lattice yields $248\,669$ candidates with non-negative expected dimension.
After imposing the boundedness criterion (that the recession cone of the sign pattern is $\{0\}$), we obtain $1\,106$ valid points, which organize into $16$ non-empty $W_\nu$-orbits.
The maximal dimension among these orbits is $2$. The final data are collected in Table~\ref{tab:e6_nu43_orbit_weight_dim}.

\begin{longtable}{c|c|c|c}
\caption{Orbit decomposition for $E_6$, slope $\nu=4/3$ and $f=\mathrm{regular}$.  Listed are the coroot-vector representatives $\beta$, the expected dimension $\dim=\mathrm{NumberA}-\mathrm{NumberB}$, and the conformal weight $h_\lambda$.}
\label{tab:e6_nu43_orbit_weight_dim}\\
\hline
Orbit & $\beta$ & $\dim$ & $h_\lambda$ \\
\hline
\endfirsthead
\hline
Orbit & $\beta$ & $\dim$ & $h_\lambda$ \\
\hline
\endhead

$\mathcal O_{1}$ & $(-13,-18,-25,-35,-25,-13)$ & $0$ & $0$ \\
$\mathcal O_{2}$ & $(-13,-18,-25,-33,-23,-12)$ & $0$ & $\frac{1}{2}$ \\
$\mathcal O_{3}$ & $(-13,-18,-24,-34,-24,-12)$ & $0$ & $-\frac{1}{2}$ \\
$\mathcal O_{4}$ & $(-13,-17,-25,-33,-23,-13)$ & $0$ & $\frac{1}{2}$ \\
$\mathcal O_{5}$ & $(-13,-17,-24,-33,-23,-13)$ & $0$ & $-\frac{3}{4}$ \\
$\mathcal O_{6}$ & $(-13,-17,-23,-31,-22,-10)$ & $0$ & $1$ \\
$\mathcal O_{7}$ & $(-12,-18,-24,-34,-24,-13)$ & $0$ & $-\frac{1}{2}$ \\
$\mathcal O_{8}$ & $(-12,-18,-23,-33,-25,-13)$ & $0$ & $\frac{1}{2}$ \\
$\mathcal O_{9}$ & $(-12,-17,-24,-33,-23,-12)$ & $0$ & $-1$ \\
$\mathcal O_{10}$ & $(-12,-17,-23,-33,-24,-12)$ & $0$ & $-1$ \\
$\mathcal O_{11}$ & $(-12,-17,-23,-33,-23,-12)$ & $1$ & $-\frac{3}{2}$ \\
$\mathcal O_{12}$ & $(-12,-17,-23,-32,-23,-12)$ & $1$ & $-2$ \\
$\mathcal O_{13}$ & $(-12,-16,-22,-31,-22,-12)$ & $2$ & $-\frac{5}{2}$ \\
$\mathcal O_{14}$ & $(-11,-16,-23,-33,-23,-11)$ & $0$ & $\frac{3}{2}$ \\
$\mathcal O_{15}$ & $(-11,-16,-21,-30,-21,-11)$ & $1$ & $-\frac{11}{4}$ \\
$\mathcal O_{16}$ & $(-11,-15,-21,-29,-21,-11)$ & $2$ & $-3$ \\
\hline

\end{longtable}

The dimension distribution of the $16$ non-empty orbits is
\begin{equation*}
\dim=0:\;11\text{ orbits},\qquad
\dim=1:\;3\text{ orbits},\qquad
\dim=2:\;2\text{ orbits}.
\end{equation*}
Comparing our results with those of \cite{jiang2024modularity,pan2025modularity}, we find agreement for the set of rational scaling dimensions; moreover, the multiplicities found there match the dimensions of our fixed varieties in the higher-dimensional cases.
The cohomology groups of the higher-dimensional fixed varieties also match the Jordan structure predicted by the character formulas of \cite{pan2025modularity}.

\subsection{\texorpdfstring{$E_8$}{E8}}

The exceptional Lie algebra $\mathfrak{g}=E_8$ admits several non-admissible
$W$-algebras whose modules can be studied via our geometric framework.
These theories admit an alternative engineering via type~IIB string theory on
hypersurface singularities, from which one can extract isomorphisms to
admissible $W$-algebras of lower rank.  The isomorphism arises because the
same singular curve admits two different decompositions into sums of
$ADE$ singularities.  For example, the singularity
\begin{equation*}
x^2+y^3+z^5+w^2=0
\end{equation*}
can be decomposed as $f_{E_8}(x,y,z)+f_{A_1}(w)=0$ or as
$f_{A_2}(x,y,w)+f_{A_4}(z)=0$, which implies the $W$-algebra isomorphism
\begin{equation*}
W^{-30+\frac{15}{16}}(E_8,f_{\mathrm{prin}})
\;\simeq\;
W^{-3+\frac{3}{8}}(A_2,f_{\mathrm{prin}}).
\end{equation*}
Our computation of the module spectrum confirms this isomorphism.
Table~\ref{e8iso} lists several such isomorphisms; each has been verified
by an explicit count of fixed varieties on the affine Springer fiber.

\begin{table}[htp]
\caption{Isomorphisms of $W$-algebras involving $E_8$.}
\begin{center}
\begin{tabular}{|c|c|} \hline
$x^2+y^3+z^5+w^2=0$ &
  $W^{-30+\frac{15}{16}}(E_8,f_{\mathrm{prin}})=
   W^{-3+\frac{3}{8}}(A_2,f_{\mathrm{prin}})$ \\ \hline
$x^2+y^3+z^5+w^3=0$ &
  $W^{-30+\frac{10}{11}}(E_8,f_{\mathrm{prin}})=
   W^{-6+\frac{6}{11}}(D_4,f_{\mathrm{prin}})$ \\ \hline
$x^2+y^3+z^5+w^4=0$ &
  $W^{-30+\frac{15}{17}}(E_8,f_{\mathrm{prin}})=
   W^{-12+\frac{12}{17}}(E_6,f_{\mathrm{prin}})$ \\ \hline
$x^2+y^3+z^5+z w=0$ &
  $W^{-30+\frac{24}{25}}(E_8,f_{\mathrm{prin}})=
   W^{-2+\frac{2}{5}}(A_1,f_{\mathrm{prin}})$ \\ \hline
$x^2+y^3+z^5+y w=0$ &
  $W^{-30+\frac{20}{21}}(E_8,f_{\mathrm{prin}})=
   W^{-2+\frac{2}{7}}(A_1,f_{\mathrm{prin}})$ \\ \hline
$x^2+y^3+z^5+y w^3=0$ &
  $W^{-30+\frac{20}{23}}(E_8,f_{\mathrm{prin}})=
   W^{-18+\frac{18}{21}}(E_7,f_{\mathrm{prin}})$ \\ \hline
\end{tabular}
\end{center}
\label{e8iso}
\end{table}

Another particularly interesting theory is the one engineered from the
singularity $x^2+y^3+z^5+w^6=0$.  This theory admits a gauge-theory
description as an $SU(5)$ gauge theory coupled to $D_2(SU(5))$, $D_3(SU(5))$, and
$D_6(SU(5))$ matter, and the corresponding $W$-algebra is
$W^{-30+\frac{6}{5}}(E_8,f_{\mathrm{prin}})$.  Using our method, we have
computed the full list of fixed varieties at this slope; the result
provides a prediction for the VOA module spectrum.  A partial list
appears in the Appendix C, and a detailed study of this model will appear
elsewhere.

\section{Twisted theories}
\label{section:twisted}
We first explain the isomorphism between the counting of the affine Springer fiber and that of the corresponding $W$-algebra for the boundary admissible case.
On the affine Springer fiber, we have a twisted affine Lie algebra with finite Lie algebra $\mathfrak{g}$. On the $W$-algebra side, we have an untwisted
affine Lie algebra with finite Lie algebra $\mathfrak{g}^\vee$, which is the Langlands dual of $\mathfrak{g}$. Choose a root system of $\mathfrak{g}$;
the simple roots are labeled as $\alpha_1,\ldots,\alpha_r$. The standard normalization of the Killing pairing is such that \textbf{long} roots have length squared $2$.
Thus the Weyl vector satisfies $\rho^2=\frac{h^\vee\dim\mathfrak{g}}{12}$.

We first focus on the Coxeter case, namely when the slope on the affine Springer fiber is $\nu=\frac{u}{h_\theta}$.
On the affine Springer fiber side, the set $S_u$ is given as
\begin{equation*}
S_u=\{\alpha_1,\ldots,\alpha_r,-\theta_s+\tfrac{u}{2}\delta\} .
\end{equation*}
where $\theta_s$ is the highest short root.
We count the affine Weyl group elements $\tilde{w}=st_q$ satisfying $\tilde{w}(S_u)\in\tilde{n}_f$, where $\tilde{n}_f$ is the subalgebra associated with $f$
(when $f=0$, $\tilde{n}_f$ consists of positive affine roots). An important subtlety is that the expansion of the translation vector $q=\sum m_i\alpha_i^\vee$ has
$m_i$ half-integers for short roots (thirds for $G_2$).

On the $W$-algebra side, we have the set $\tilde{S}_u$, which involves the coroots of $\mathfrak{g}^\vee$:
\begin{equation*}
\tilde{S}_u=\{\alpha_1^{\prime\vee},\ldots,\alpha_r^{\prime\vee},-\theta_l^{\prime\vee}+u\delta\}
=\{\alpha_1,\ldots,\alpha_r,-2\theta_s+u\delta\},
\end{equation*}
where $\alpha_i^{\prime}$ are the simple roots of $\mathfrak{g}^\vee$. By Langlands duality, $\alpha_i^{\prime\vee}=\alpha_i$ and $\theta_l^{\prime\vee}=2\theta_s$.
Note that with this identification, the inner product on the pairing of $\mathfrak{g}^\vee$ is no longer standard, since now the \textbf{short} root has length squared $2$.
Using the standard normalization, we get the same set as that of the affine Springer fiber.

The condition for finding irreducible modules also involves the extended affine Weyl group element of the untwisted Lie algebra (modulo an equivalence condition) \cite{kac2008rationality}.
The affine Weyl group element is $st_q$, where $q$ lies in the coweight lattice of $\mathfrak{g}^\vee$, which contains the root lattice of $\mathfrak{g}$.
The condition is the same: $\tilde{w}(\tilde{S}_u)\in\tilde{n}_f$. We have thus proved the equivalence of the two counting problems.

Because of the above subtle normalization of the Killing pairing, one must be careful when computing scaling dimensions of modules from affine Springer fiber data.
We explain this in detail:
\begin{enumerate}
\item Start with the twisted affine Springer fiber and choose the normalization of the finite Lie algebra $\mathfrak{g}$
so that the \textbf{short} root has length squared $2$ (so the coroots which are the root of dual algebra $\mathfrak{g}^\vee$ has the standard normalization.). Compute the affine Weyl group element $\tilde{w}=st_q$ satisfying the condition \eqref{equ1} and \eqref{equ2} (and check the Chern constraint at the end).
Here $q$ lives in the coroot lattice of $\mathfrak{g}$, and the coefficient before short coroots takes half-integer values (thirds for $G_2$).
\item Next, pass to the $W$-algebra where we have the Lie algebra $\mathfrak{g}^\vee$. Due to the normalization, for an affine Weyl group element $st_q$ computed from the affine Springer fiber side,
the affine Weyl element on the $W$-algebra side is $st_{q'}$, with
\begin{equation*}
q' = nq
\end{equation*}
where $n=2$ for all twisted cases except twisted $G_2$ (where $n=3$).
We then obtain a finite weight by computing $\bar{\lambda}=st_{q'}(k\Lambda_0+\tilde{\rho})-\tilde{\rho}$. When $s=1$, the end result is simply a rescaling of $q$:
\begin{equation*}
\bar{\lambda}=(k+h^\vee)q' = \frac{1}{n}\frac{m}{u}q' = \frac{m}{u} q
\end{equation*}
Therefore the final finite vector takes the same form for the twisted and untwisted theories: multiply the slope of the affine Springer fiber.
\item Compute the scaling dimension of the highest-weight module $M(\bar{\lambda})$ of the $W$-algebra using the formula \eqref{scale}, and the $W$-algebra is defined using the Langlands dual Lie algebra $\mathfrak{g}^\vee$.
The Weyl vector $\rho_{\mathfrak{g^\vee}}$ can be expressed in terms of the positive coroots of $\mathfrak{g}$. The dual Weyl vector $\rho_{\mathfrak{g}^\vee}^\vee$ can also be computed using the positive roots of $\mathfrak{g}$.  The pairing in computing the scaling dimension
is the one in the first step of the pipeline.
\end{enumerate}

\subsection{\texorpdfstring{$^{3}D_4$}{3D4}}

\subsubsection{\texorpdfstring{$m=12$}{m=12}}
For slope $\nu=\frac{u}{12}$, the affine Springer fiber has only zero-dimensional fixed varieties. The count is
\begin{equation*}
N_0(u)=\frac{(u-1)(u-5)}{8}.
\end{equation*}
We now explain some details of the computation of the fixed varieties.
Take the simple roots of $G_2$ as $\alpha_1,\alpha_2$, with $\alpha_1$ long and $\alpha_2$ short. The pairing is $(\alpha_1,\alpha_1)=6,\;(\alpha_1,\alpha_2)=-3,\;(\alpha_2,\alpha_2)=2$.
The set of short positive roots is $\Phi^0_{s+}=\{\alpha_2, \alpha_1+\alpha_2, \alpha_1+2\alpha_2\}$, and the set of positive long roots is $\Phi^0_{l+}=\{\alpha_1, \alpha_1+3\alpha_2, 2\alpha_1+3\alpha_2\}$.

The roots of the twisted affine Lie algebra are
\begin{align}
    & \Phi^{re}_{s}=\{\alpha+\frac{n}{3}\delta ~|~ \alpha \in \Phi^0_s ,~n\in\bbZ\}, \nonumber\\
    &  \Phi^{re}_{l}=\{\alpha+n\delta~|~ \alpha \in \Phi^0_l,~n\in\bbZ \}.
\end{align}
The set $L_u$ is empty, and the set $S_u$ consists of roots
\begin{equation*}
S_u=\{\alpha_1, \alpha_2, -\theta_s+\frac{u}{3}\delta\}
\end{equation*}
Here $\theta_s=\alpha_1+2\alpha_2$ is the highest short root.
The affine Weyl group of the twisted theory is written as $st_q$, where $q=m_1\alpha_1^\vee+m_2\alpha_2^\vee$, and
$m_2$ takes values in thirds. The solutions for the fixed points are
labeled by affine Weyl group elements $\tilde{w}$ such that $\tilde{w}(S_u)\in\tilde{n}_f$. For example, for $f=\mathrm{regular}$,
the affine Weyl group element is $\tilde{w}=t_{m_1\alpha_1^\vee+m_2\alpha_2^\vee}$, and the conditions on the integers $m_1$ and $m_2$ are
\begin{equation}
\begin{cases}
3m_2>2m_1,\\[2pt]
3m_1>6m_2,\\[2pt]
u+3m_2>0
\end{cases}
\label{affine}
\end{equation}
We note that $m_2$ takes values in thirds.

For $u=11$ and $f=\mathrm{regular}$, the Coulomb branch spectrum suggests an isomorphism with the $(2,11)$ Virasoro minimal model.
There are five solutions $t_q$ from the affine Springer fiber side, namely
\begin{equation}
(m_1,3m_2)=(-6,-10),(-5,-9),(-5,-8),(-4,-7),(-3,-5),
\end{equation}
where $q=m_1\alpha_1^\vee+m_2\alpha_2^\vee$.

The Weyl vector $\rho$ on the $W$-algebra side is $\rho=3\alpha_l^\vee+5\alpha_s^\vee=\frac{5}{3}\alpha_1+5\alpha_2$ and $\rho^\vee=3\alpha_1+5\alpha_2$.
Computing the scaling dimensions using the formula \eqref{scale} (with level $k=-4+\frac{4}{11}$ and finite vector
$\bar{\gamma}=\frac{12}{11}q$) gives
$(-\frac{4}{11},-\frac{10}{11},-\frac{7}{11},-\frac{9}{11},0)$,
which exactly match those of the $(2,11)$ Virasoro minimal model.

\subsubsection{\texorpdfstring{$m=6$}{m=6}}
Assume the slope is
\[
\nu=\frac{u}{6}, \qquad \gcd(u,6)=1,
\]
for the twisted setting \(^{3}D_{4}\to G_{2}\), with \(\alpha_{1}\) long and \(\alpha_{2}\) short.
The level quantization is:
\[
k\in \mathbb{Z}\ \text{for long roots},
\qquad
k\in \tfrac{1}{3}\mathbb{Z}\ \text{for short roots}.
\]
Define
\[
L_{\nu}:=\{(\alpha,k)\mid u(\alpha,\rho^\vee)+6k=0\},
\qquad
S_{\nu}:=\{(\alpha,k)\mid u(\alpha,\rho^\vee)+6k=u\}.
\]
Then one gets
\[
L_{\nu}
=
\left\{
\left(\alpha_{1}+\alpha_{2},-\frac{u}{3}\right),
\left(-\alpha_{1}-\alpha_{2},\frac{u}{3}\right)
\right\},
\]
and
\[
\begin{aligned}
S_{\nu}
=
\Bigl\{&
(\alpha_{1},0),\;
(-2\alpha_{1}-3\alpha_{2},u),\;
(\alpha_{2},0),\;
\left(\alpha_{1}+2\alpha_{2},-\frac{u}{3}\right),\\
&\left(-\alpha_{2},\frac{u}{3}\right),\;
\left(-\alpha_{1}-2\alpha_{2},\frac{2u}{3}\right)
\Bigr\}.
\end{aligned}
\]
Hence
\[
|L_{\nu}|=2,\qquad |S_{\nu}|=6.
\]

For slope $\nu=\frac{u}{6}$, the counting is
\[
N_{\mathrm{orb}}^{(1)}(u)=\left\lfloor \frac{(n-1)^2}{4}\right\rfloor
=\left\lfloor \frac{(u-3)^2}{16}\right\rfloor,
\]
\[
N_{\mathrm{orb}}^{(0)}(u)=N_{\mathrm{orb}}(u)-N_{\mathrm{orb}}^{(1)}(u)
=\left(\frac{u-1}{2}\right)^2-\left\lfloor \frac{(u-3)^2}{16}\right\rfloor.
\]
There are only zero-dimensional and one-dimensional varieties. Some concrete examples are listed in Table~\ref{tab:g2_u5_u7_orbits}.
\begin{table}[htbp]
    \centering
    \caption{Orbit representatives of admissible $\beta=m_1\alpha_1^\vee+m_2\alpha_2^\vee$ for $\nu=u/6$ ($u=5,7$), under the right action generated by affine roots in $L_\nu$.
    We record $(m_1,\,3m_2)$ (so all entries are integers), the dimension $\dim:=\mathrm{NumberA}-\mathrm{NumberB}$, and the  scaling dimension $h_\lambda$.}
    \label{tab:g2_u5_u7_orbits}
    \begin{tabular}{c c c c c}
    \hline
    $u$ & Orbit rep.\ index & $(m_1,\,3m_2)$ & $\dim$ & $h_\lambda$ \\
    \hline
    5 & 1 & $(-5,-9)$  & 0 & $-\frac{2}{5}$ \\
    5 & 2 & $(-5,-8)$  & 0 & $-\frac{1}{5}$ \\
    5 & 3 & $(-4,-7)$  & 0 & $-\frac{3}{5}$ \\
    5 & 4 & $(-4,-6)$  & 0 & $0$ \\
    \hline
    7 & 1 & $(-8,-13)$ & 0 & $-1$ \\
    7 & 2 & $(-7,-13)$ & 0 & $-\frac{12}{7}$ \\
    7 & 3 & $(-7,-12)$ & 0 & $-\frac{13}{7}$ \\
    7 & 4 & $(-7,-11)$ & 0 & $-\frac{10}{7}$ \\
    7 & 5 & $(-6,-11)$ & 0 & $-\frac{15}{7}$ \\
    7 & 6 & $(-6,-10)$ & 1 & $-2$ \\
    7 & 7 & $(-6,-9)$  & 0 & $-\frac{9}{7}$ \\
    7 & 8 & $(-6,-8)$  & 0 & $0$ \\
    7 & 9 & $(-5,-8)$  & 0 & $-\frac{11}{7}$ \\
    \hline
    \end{tabular}
    \end{table}

For $u=5$ and $f=\mathrm{regular}$, the Coulomb branch spectrum has scaling dimension $(\frac{6}{5},\frac{6}{5},\frac{6}{5})$. By computing the 4d central charges,
it is predicted that the low-energy theory consists of three identical copies of the $(A_1,A_2)$ theory (which has only one operator of dimension $\frac{6}{5}$).
The corresponding VOA is therefore the tensor product of three copies of the $(2,5)$ minimal model.
There are four irreducible modules with scaling dimensions $(0,-\frac{1}{5},-\frac{2}{5},-\frac{3}{5})$ (the multiplicities are computed by counting the points in each fixed variety and are $(1,3,3,1)$), which agrees with the tensor product of the modules of the $(2,5)$ model
(which has two irreducible modules: the vacuum module of scaling dimension zero, and one non-trivial module of dimension $-\frac{1}{5}$). The fixed varieties for $u=5,7$ are listed in Table~\ref{tab:g2_u5_u7_orbits}.

\subsubsection{\texorpdfstring{$m=3$}{m=3}}
The set
\[
\begin{aligned}
L_{\nu}
=\Bigl\{&
\left(\alpha_{2},-\frac{u}{3}\right),
\left(-\alpha_{2},\frac{u}{3}\right),
\left(\alpha_{1}+\alpha_{2},-\frac{2u}{3}\right),
\left(-\alpha_{1}-\alpha_{2},\frac{2u}{3}\right),\\
&(\alpha_{1}+2\alpha_{2},-u),
(-\alpha_{1}-2\alpha_{2},u)
\Bigr\},
\end{aligned}
\]
and
\[
\begin{aligned}
S_{\nu}
=\Bigl\{&
(\alpha_{1},0),
(\alpha_{2},0),
\left(-\alpha_{2},\frac{2u}{3}\right),
\left(\alpha_{1}+\alpha_{2},-\frac{u}{3}\right),
(-\alpha_{1}-\alpha_{2},u),\\
&\left(\alpha_{1}+2\alpha_{2},-\frac{2u}{3}\right),
\left(-\alpha_{1}-2\alpha_{2},\frac{4u}{3}\right),
(\alpha_{1}+3\alpha_{2},-u),
(-2\alpha_{1}-3\alpha_{2},2u)
\Bigr\}.
\end{aligned}
\]
In particular,
\[
|L_{\nu}|=6,\qquad |S_{\nu}|=9.
\]

For slope $\nu=\frac{u}{3}$, the counting of the fixed varieties is as follows.
\begin{table}[htbp]
  \centering
  \scriptsize
  \setlength{\tabcolsep}{2pt}
  \caption{Quadratic formulas for orbit counts at slope $\nu=\frac{u}{3}$ ($u\perp 3$). The formulas depend on $u$ modulo $6$.}
  \label{tab:g2-u-over-3-quadratic-formulas}
  \begin{tabular}{c|c c c c|c}
    \hline
    Residue class & $N_0(u)$ & $N_1(u)$ & $N_2(u)$ & $N_3(u)$ & $N_{\mathrm{orb}}(u)$ \\
    \hline
    $u\equiv 1 \pmod 6$ &
    $\dfrac{u(u-1)}{6}$ &
    $\dfrac{(u-1)(u+1)}{6}$ &
    $\dfrac{(u-1)(u+7)}{12}$ &
    $\dfrac{(u-5)(u-1)}{12}$ &
    $\dfrac{(u-1)(3u+2)}{6}$ \\[8pt]
    $u\equiv 2 \pmod 6$ &
    $\dfrac{u^2-u+4}{6}$ &
    $\dfrac{u^2+2}{6}$ &
    $\dfrac{(u-2)(u+8)}{12}$ &
    $\dfrac{(u-4)(u-2)}{12}$ &
    $\dfrac{3u^2-u+2}{6}$ \\[8pt]
    $u\equiv 4 \pmod 6$ &
    $\dfrac{u(u-1)}{6}$ &
    $\dfrac{u^2+2}{6}$ &
    $\dfrac{(u-2)(u+8)}{12}$ &
    $\dfrac{(u-4)(u-2)}{12}$ &
    $\dfrac{(u-1)(3u+2)}{6}$ \\[8pt]
    $u\equiv 5 \pmod 6$ &
    $\dfrac{u^2-u+4}{6}$ &
    $\dfrac{(u-1)(u+1)}{6}$ &
    $\dfrac{(u-1)(u+7)}{12}$ &
    $\dfrac{(u-5)(u-1)}{12}$ &
    $\dfrac{3u^2-u+2}{6}$ \\
    \hline
  \end{tabular}
\end{table}

When $u=2$, computing the central charges $(a,c)=(5/24,1/6)$ shows that the 4d theory is a free vector multiplet.
The 2d theory has central charge $c=-2$, and the VOA is the symplectic fermion VOA.
The corresponding $W$-algebra is $W^{-4+\frac{1}{2}}(\mathfrak{g}_2,f_{\mathrm{prin}})$.
On the affine Springer fiber side, the slope is $\frac{3}{2}$.
There is one zero-dimensional variety and one one-dimensional variety. Using the root data in Table~\ref{tab:g2-u2-orbit-dim} and
the formula for scaling dimension \eqref{scale}, we obtain $h=0$ for the one-dimensional variety and $h=1$ for the zero-dimensional variety.
Our computation predicts that the VOA has two simple modules with scaling dimensions $0$ and $1$, together with a logarithmic module.
Indeed, for the $c=-2$ triplet $W$-algebra, the bosonic sector consists of two simple modules $W_0, W_1$ with the same scaling dimension.
The $c=-2$ model studied in \cite{flohr1996modular} contains other modules. We leave the study of the modular properties of the present model to future work.

 \begin{table}[htbp]
  \centering
  \caption{Right-action orbit representatives for twisted $G_2$ at $\nu=\frac{2}{3}$, together with their raw scaling dimensions $h_\lambda$.}
  \label{tab:g2-u2-orbit-dim}
  \begin{tabular}{c c c c}
    \hline
    Orbit & Representative $(m_1,3m_2)$ & $\dim=\mathrm{NumberA}-\mathrm{NumberB}$ & $h_\lambda$ \\
    \hline
    $\mathcal O_1$ & $(-4,-7)$ & $1$ & $0$ \\
    $\mathcal O_2$ & $(-3,-7)$ & $0$ & $1$ \\
    \hline
  \end{tabular}
\end{table}

When $u=4$, the theory has the same Coulomb branch spectrum as the theory
 engineered by the threefold singularity $x^2+y^3+z^4+w^4=0$.
The fixed points and fixed varieties are listed in Table~\ref{tab:g2-u4-orbit-dim}. Comparing the results with those using $E_6$ data (see Table~\ref{tab:e6_nu43_orbit_weight_dim}), we see that the higher-dimensional varieties agree
with each other, and it would be interesting to explore the difference for the other modules.

\begin{table}[htbp]
  \centering
  \caption{Right-action orbit representatives for twisted $G_2$ at $\nu=\frac{4}{3}$, together with their raw scaling dimensions $h_\lambda$.}
  \label{tab:g2-u4-orbit-dim}
  \begin{tabular}{c c c c}
    \hline
    Orbit & Representative $(m_1,3m_2)$ & $\dim$ & $h_\lambda$ \\
    \hline
    $\mathcal O_1$ & $(-10,-19)$ & $0$ & $0$ \\
    $\mathcal O_2$ & $(-9,-17)$  & $1$ & $-\frac{3}{2}$ \\
    $\mathcal O_3$ & $(-9,-16)$  & $1$ & $-2$ \\
    $\mathcal O_4$ & $(-8,-15)$  & $2$ & $-\frac{5}{2}$ \\
    $\mathcal O_5$ & $(-8,-14)$  & $1$ & $-\frac{11}{4}$ \\
    $\mathcal O_6$ & $(-7,-13)$  & $2$ & $-3$ \\
    $\mathcal O_7$ & $(-6,-14)$  & $0$ & $-\frac{3}{4}$ \\
    \hline
  \end{tabular}
\end{table}

\subsection{\texorpdfstring{Twisted $^2A_3$}{Twisted 2A3}}

\subsubsection{\texorpdfstring{$m=6$}{m=6}}
For the twisted $^2A_3$ theory, the only allowed values are $m=6,2$. The value $m=6$ is the Coxeter one.
This is the regular elliptic case. The root system is taken to be $\alpha_1,\alpha_2$ with $\alpha_1$ a long root and $\alpha_2$ a short root. The pairing is
$(\alpha_1,\alpha_1)=4,\;(\alpha_1,\alpha_2)=-2,\;(\alpha_2,\alpha_2)=2$. The short highest root is $\theta_s=\alpha_1+\alpha_2$, while
the long highest root is $\theta_l=\alpha_1+2\alpha_2$.

The affine Weyl group of the twisted Lie algebra takes the form $st_q$, where $q$ belongs to the coroot lattice.
The set $L_\nu$ is empty.
Fixed points are determined from $S_\nu$ at slope $\nu=\frac{u}{6}$, where
\begin{equation*}
S_{u/6}=\{\alpha_1,\alpha_2,-\theta_s+\tfrac{u}{2}\delta\} .
\end{equation*}

In particular, when $f$ is regular (so we obtain the principal $W$-algebra), the fixed varieties are counted as follows.
Take the translation $t_q=t_{m_1\alpha_1^\vee+m_2\alpha_2^\vee}$;
the condition $t_q(S_u)\in\tilde{\Delta}_+/\Delta_+$ becomes
\begin{equation*}
\begin{cases}
2m_1-2m_2 < 0,\\[2pt]
-m_1+2m_2 < 0,\\[2pt]
\displaystyle\frac{u}{2}+m_1 > 0
\end{cases}
\end{equation*}
We emphasize that $m_2$ can take half-integer values. The number of zero-dimensional varieties (and hence the number of irreducible modules on the VOA) is given by
\begin{equation*}
N_0=\frac{(u-1)(u-3)}{8}.
\end{equation*}

The $W$-algebra under consideration is
\begin{equation*}
W^k(\mathfrak{B}_2,f_{\mathrm{prin}}),\qquad k=-3+\frac{3}{u},\qquad (u,3)=1.
\end{equation*}

When $u=7$, the theory is isomorphic to the $(A_1,A_4)$ theory, whose corresponding VOA is the Virasoro $(2,7)$ minimal model, which should have three modules.
On the affine Springer fiber side, there are three fixed points with affine translation $t_q$ and $q=(m_1,2m_2)=(-3,-5),(-3,-4),(-2,-3)$ in the basis of $\alpha_1,\alpha_2$.
The finite weight on the $W$-algebra side is $\frac{6}{7}q$.
The Weyl vector of the $B_2$ Lie algebra is $\rho_{B_2}=\alpha_1+\frac{3}{2}\alpha_2$; and the dual Weyl vector $\rho^\vee_{B_2}=\frac{3}{2}\alpha_1+2\alpha_2$ (half of the positive roots of $C_2$ Lie algebra).
We also have $(\rho^\vee)^2=5$ and $(\rho,\rho^\vee)=\frac{7}{2}$.  For the three solutions we find the scaling dimensions $(0,-\frac{2}{7},-\frac{3}{7})$, which match exactly the irreducible modules of the Virasoro minimal model $(2,7)$.

\subsubsection{\texorpdfstring{$m=2$}{m=2}}
The $W$-algebra we consider is
\begin{equation*}
W^k(\mathfrak{B}_2, f),\qquad k=-3+\frac{1}{u},
\end{equation*}
where $u$ is an odd positive integer.

\paragraph{Twisted type $^{2}A_{3}$ at slope $\nu=\frac{u}{2}$ (with $\gcd(u,2)=1$).}
Let the finite root system be of type $B_{2}/C_{2}$ with simple roots
$\alpha_{1}$ (long) and $\alpha_{2}$ (short), and positive roots
\[
\Phi^{+}=\{\alpha_{1},\,\alpha_{2},\,\alpha_{1}+\alpha_{2},\,\alpha_{1}+2\alpha_{2}\}.
\]
Using $(\alpha_i,\rho^\vee)=1$, we have
\[
(\,n_{1}\alpha_{1}+n_{2}\alpha_{2},\,\rho^\vee)=n_{1}+n_{2}.
\]
For affine roots $(\alpha,k)$, take
\[
k\in\mathbb Z \ \text{for long roots},
\qquad
k\in \tfrac12\mathbb Z \ \text{for short roots}.
\]
Define
\[
L_\nu=\{(\alpha,k)\mid u(\alpha,\rho^\vee)+2k=0\},
\qquad
S_\nu=\{(\alpha,k)\mid u(\alpha,\rho^\vee)+2k=u\}.
\]
Then (for odd $u$):
\begin{align*}
L_\nu
=\Big\{&
(\alpha_{2},-\tfrac u2),\,
(-\alpha_{2},\tfrac u2),\,
(\alpha_{1}+\alpha_{2},-u),\,
(-(\alpha_{1}+\alpha_{2}),u)
\Big\},
\\[2mm]
S_\nu
=\Big\{&
(\alpha_{1},0),\,
(-\alpha_{1},u),\,
(\alpha_{1}+2\alpha_{2},-u),\,
(-(\alpha_{1}+2\alpha_{2}),2u),\\
&(\alpha_{2},0),\,
(-\alpha_{2},u),\,
(\alpha_{1}+\alpha_{2},-\tfrac u2),\,
(-(\alpha_{1}+\alpha_{2}),\tfrac{3u}{2})
\Big\}.
\end{align*}

We now examine the dimension formula for the Hessenberg variety; see equation~\eqref{eq:hessenberg-dim}. Since $\mathrm{NumberA}$ is always 2, and $\mathrm{NumberB}$ is at most 2,
there are at most two elements in $S_u$ that
would have negative signs for the region $\tilde{w}^{-1}(\Delta_0)$. By looking at the set $S_u$, it is easy to see that
any choice of two signs gives a bounded region. Therefore, we only need to compute $\tilde{w}$ that gives
non-negative expected dimension of the Hessenberg variety.

When $f=0,\;\nu=\frac12$, the fixed varieties are
found in \cite{oblomkov2016geometric}: there is one two-dimensional fixed variety, namely $\mathbb P^1\times \mathbb P^1$.
There is one one-dimensional fixed variety, which is an elliptic curve $C$, and two zero-dimensional fixed varieties. The $S_a\times B_a$-invariant
part of the cohomology is $H^*(\mathbb P^1\times \mathbb P^1)\oplus H^*(pt)\oplus H^*(pt)\oplus H^*(C)/H^1(C)$. One needs to remove the degree-one cohomology
of the torus. Therefore the total dimension is $8$. For general $u$, the number of solutions for $\tilde{w}$ giving non-empty cells via~\eqref{eq:cell-nonempty} is $16u^2$.
After quotienting by the $\bbZ_2\times\bbZ_2$ symmetry generated by $L_u$, the number of independent solutions is $4u^2$: $2u^2$ two-dimensional varieties, $u^2$ one-dimensional varieties, and
$2u^2$ zero-dimensional varieties.

We now consider $u=3$ and $f=\mathrm{regular}$, so that we obtain the principal $W$-algebra. The Coulomb branch spectrum of this theory is $(2,\frac{4}{3},\frac{4}{3},\frac{4}{3})$,
thus the physical theory also admits a gauge-theory description as $SU(2)$ coupled with three $D_3(SU(2))$ theories. This theory can be engineered using the threefold singularity $x^3+y^3+z^3+w^2=0$,
which was already discussed in the previous section (see Table~\ref{tab:twistedA3_u3_orbit_reps_m1_2m2} for results).
For general $u$, the numbers of varieties in various dimensions are given by
\[
N_2(u)=\frac{(u-1)(u-3)}{8},\qquad
N_1(u)=\frac{(u-1)(u+5)}{8},\qquad
N_0(u)=\frac{(u-1)^2}{4},
\]
\[
N_{\mathrm{tot}}(u)=N_2(u)+N_1(u)+N_0(u)=\frac{u(u-1)}{2}.
\]

\begin{table}[htbp]
    \centering
    \begin{tabular}{c|c|c}
    \hline
    Representative $(m_1,\,2m_2)$ & Dimension $\mathrm{NumberA}-\mathrm{NumberB}$& $h_\lambda$\\
    \hline
    $(-4,-7)$ & $0$ & 0\\
    $(-4,-6)$ & $1$ &$-\frac{2}{3}$\\
    $(-3,-5)$ & $1$ &$-1$\\
    \hline
    \end{tabular}
    \caption{Orbit representatives for twisted $^2A_3$ at slope $\nu=\frac{3}{2}$, with
    $\beta=m_1\alpha_1^\vee+m_2\alpha_2^\vee$ ($m_1\in\mathbb Z,\ m_2\in\frac12\mathbb Z$).}
    \label{tab:twistedA3_u3_orbit_reps_m1_2m2}
    \end{table}

    \begin{table}[htbp]
        \centering
        \begin{tabular}{c|c|c}
        \hline
        Representative $(m_1,\,2m_2)$ & Dimension $\mathrm{NumberA}-\mathrm{NumberB}$ & $h_\lambda$\\
        \hline
        $(-8,-12)$ & $0$ & $-1$\\
        $(-7,-13)$ & $0$ & $0$\\
        $(-7,-12)$ & $0$ & $-\frac{6}{5}$\\
        $(-7,-11)$ & $1$& $-2$\\
        $(-7,-10)$ & $1$& $-\frac{12}{5}$\\
        $(-6,-11)$ & $0$&$-\frac{9}{5}$\\
        $(-6,-10)$ & $1$& $-\frac{13}{5}$\\
        $(-6,-9)$  & $2$&$-3$\\
        $(-5,-9)$  & $1$&$-\frac{14}{5}$\\
        $(-5,-8)$  & $1$&$-\frac{16}{5}$\\
        \hline
        \end{tabular}
        \caption{Orbit representatives for twisted $^2A_3$ at slope $\nu=\frac{5}{2}$, with
        $\beta=m_1\alpha_1^\vee+m_2\alpha_2^\vee$ ($m_1\in\mathbb Z,\ m_2\in\frac12\mathbb Z$), grouped by right action generated by $L_\nu$.}
        \label{tab:twistedA3_u5_orbit_reps_m1_2m2}
        \end{table}

\newpage

\section{Conclusion}

In this paper we have established a detailed correspondence between
the representation theory of non-admissible $W$-algebras and the geometry of
generalized affine Springer fibers. The central proposal is that for a
$W$-algebra
\begin{equation*}
W^{k}(\mathfrak g,f),\qquad
k=-h^\vee+\frac{1}{n}\frac{m}{u},
\end{equation*}
the representation-theoretic data can be read off from the
$\bbC^*$-fixed loci of the affine Springer fiber
$Sp_{\nu}(\tilde{\mathfrak g},f)$ with slope $\nu=u/m$.

Concretely, each non-empty cell labeled by an affine Weyl-group element
$\tilde w$ gives rise to simple modules via the map
\begin{equation*}
\boxed{\tilde w\;\longmapsto\;\tilde w(k\Lambda_0+\tilde\rho)-\tilde\rho},
\end{equation*}
which determines the highest weight and conformal dimension.
The dimension of the associated Hessenberg variety $S_0^{\tilde w}$
controls additional non-semisimple structure:
zero-dimensional fixed varieties correspond to ordinary simple modules,
while higher-dimensional fixed varieties signal the presence of
logarithmic modules.
We tested the proposal by computing cases that are isomorphic to known admissible $W$-algebras.

\subsection*{Future directions}

\textbf{Characters and modular invariance.}
The most immediate next step is to extract the modular data of the
$W$-algebra from the geometry.
As argued in~\cite{shan2026mirror,shan2024modularity}, the cohomology of the affine Springer fiber carries an action of the double affine Hecke algebra (DAHA), which endows the space of generalized characters with a canonical $SL(2,\bbZ)$ representation.
In Part~II we will compute the resulting $S$ and $T$ matrices
explicitly for several non-admissible families and compare them with
modular differential equations and Zhu's theory~\cite{zhu1996modular}.
A key test will be whether the Verlinde formula, suitably generalized
to include logarithmic modules, produces sensible fusion rules.

\textbf{Categorical aspects.}
Beyond characters, the module category of a $C_2$-cofinite VOA is
expected to carry the structure of a \emph{braided tensor category}
(for a systematic treatment, see e.g.\ the Huang--Lepowsky--Zhang tensor
product theory \cite{huang2014logarithmic}).
On the geometric side, the affine Springer fiber is expected to give
rise to a category of coherent sheaves or perverse sheaves with a
braided monoidal structure induced by convolution.
Establishing a direct equivalence between the VOA module category and
the geometric category would lift our correspondence from the level of
characters to the level of tensor structures, providing a far-reaching
generalization of the geometric Langlands program to non-rational
settings.

\textbf{Extension to non-principal nilpotent orbits.}
In this paper we focused mainly on $f=0$ (affine Kac--Moody) and
$f=\text{regular}$ (principal $W$-algebra).
For intermediate nilpotent orbits the associated parabolic subgroup
$P$ is non-trivial, and the fixed-variety dimensions can exhibit a
richer pattern.
A systematic study of these cases---both for classical and exceptional
algebras---would substantially enlarge the class of $C_2$-cofinite
VOAs whose representation theory is accessible by geometric methods.

\textbf{Chern-class constraints.}
As discussed in Section~\ref{subsec:recap}, the boundedness condition
is necessary but not always sufficient for a cell to be non-empty;
there is an additional Chern-class constraint coming from the cohomology
of the flag manifold.
It remains to understand the precise conditions under which it is
non-trivial, and to develop efficient computational tools for
eliminating empty cells.

\textbf{Hessenberg varieties and detailed logarithmic structure.}
While the dimension of the fixed variety $S_0^{\tilde w}$ determines the
number of logarithmic modules it contributes---one for each unit of dimension---a complete understanding of these logarithmic modules---
including their fusion rules, modular transformation properties, and
embedding diagrams---requires a finer study of the topology of the
associated Hessenberg variety itself.

\newpage
\appendix

\section{Conventions}
Let $\mathfrak{g}$ be a simple Lie algebra of rank $r$, with Cartan subalgebra $\mathfrak h$ and dual $\mathfrak h^{*}$.
One can use the Killing form to identify $\mathfrak h$ and $\mathfrak h^*$.
We use $(,)$ to denote the Killing pairing.
The simple roots are denoted by $\alpha_1,\ldots, \alpha_r$, and the coroots are denoted by $\alpha_1^\vee,\ldots, \alpha_r^\vee$ (the coroot is defined as $\alpha^\vee=\frac{2\alpha}{(\alpha,\alpha)}$). The fundamental weights $w_i$ satisfy
\begin{equation*}
(w_i, \alpha_j^\vee)=\delta_{ij}
\end{equation*}
Using the fundamental weights, the Killing pairing is given by
\begin{equation*}
(\lambda, \mu)=\sum_{ij}\lambda_i\mu_jF_{ij},\qquad
F_{ij}=(A^{-1}_{ij})\frac{\alpha_j^2}{2}.
\end{equation*}
where $A_{ij}$ is the Cartan matrix, and $\lambda_i, \mu_i$ are the coefficients in the basis of fundamental weights.

The Weyl group action on a vector in the root space is generated by a simple root:
\begin{equation*}
s_\alpha(\beta)=\beta-(\beta,\alpha^\vee)\alpha .
\end{equation*}
The Weyl vector is denoted by $\rho$ and is given by the half-sum of the positive roots.

The space of affine roots is denoted by $\alpha+k\delta$, where $\alpha$ runs over the roots of the finite Lie algebra and $\delta$ is the imaginary root.
Sometimes we use the notation $(\alpha, k)$ to denote an affine root.

The affine Weyl group action of an element $\tilde{w}=st_q$ on an affine root $\beta+k\delta$ is
\begin{equation*}
st_q(\beta+k\delta)=s(\beta)+(k-(q,\beta)) \delta
\end{equation*}
The affine fundamental weights $\tilde{w}_i$ are related to the finite fundamental weights $w_i$ as follows
\begin{equation*}
 \tilde{w}_i=a_i^\vee \tilde{w}_0+w_i
\end{equation*}
where $a_i^\vee$ is the comark of the finite Lie algebra.
The level of a finite fundamental weight is zero, while the level of $\tilde{w}_0$ is one. For an affine weight with level $k$, one can
either express it in terms of the affine fundamental weights, or simply list the combination of finite fundamental weights (the coefficient of $\tilde{w}_0$ equals the level $k$).
The affine Weyl vector $\tilde{\rho}$ has coordinates $[1,\ldots,1]$ in the basis of affine fundamental weights.

\section{Boundedness condition}
We outline the computational procedure used to determine the fixed loci and check the boundedness criterion~\eqref{eq:cell-nonempty}.

For an affine Weyl element $\tilde w$, pick an interior point of its alcove:
\[
x_{\tilde w}=\tilde w\cdot x_0,\qquad x_0\in\Delta^\circ .
\]
The sign pattern (clan) of $x_{\tilde w}$ with respect to the $\nu$-walls is then read off.

\medskip\noindent\textbf{Boundedness criterion.}
Write the $\nu$-walls as affine hyperplanes
\[
H_i=\{x\mid \beta_i(x)+k_i=0\},\quad i=1,\dots,N,
\]
with $\beta_i$ the root part (linear). The clan containing $x_{\tilde w}$ is determined by the signs
\[
\sigma_i=\operatorname{sgn}\bigl(\beta_i(x_{\tilde w})+k_i\bigr)\in\{\pm1\}.
\]

The corresponding region is
\[
C_\sigma=\bigl\{x\mid \sigma_i(\beta_i(x)+k_i)>0\ \forall i,\ \text{(dominant constraints)}\bigr\}.
\]
Boundedness is equivalent to the recession cone being trivial:
\[
\operatorname{rec}(C_\sigma)
=\bigl\{v\mid \sigma_i\,\beta_i(v)\ge 0\ \forall i,\ \text{(dominant linear constraints on }v)\bigr\}
=\{0\}.
\]
Equivalently,
\[
C_\sigma\text{ bounded}
\iff
\nexists\,v\neq 0\text{ such that }
\begin{cases}
\sigma_i\,\beta_i(v)\ge 0,\ \forall i,\\
\text{(dominant linear constraints)}.
\end{cases}
\]
A computationally convenient form is:
\[
\exists v\neq0\text{ with those inequalities}
\iff
\exists v\text{ with those inequalities and }\|v\|^2\ge 1.
\]
If the system is feasible the region is unbounded; if infeasible it is bounded.

Now take $x_0=c\rho_P^\vee$ in the interior of the fundamental alcove, with $c$ a sufficiently small positive number. Given an affine Weyl group element $\tilde{w}=st_q$, its inverse is $\tilde{w}^{-1}=t_{-q}s^{-1}$,
and its action on $x_0$ is $\tilde{w}^{-1}(x_0)=s^{-1}(x_0)-q$. Its sign for the hyperplane defined by the element $(\alpha,k)$ is
\begin{equation*}
(\alpha,s^{-1}(c\rho_P^\vee)-q)+k=(s(\alpha),c\rho_P^\vee)-(q,\alpha)+k .
\end{equation*}
This is exactly the condition studied in formula~\eqref{equ2}, which defines the root space $V_{\tilde{y},t}$. With $t>0$, the denominator of the second term in~\eqref{equ2} gives
the number of positive signatures. From the above computation, the other roots in $S_\nu$ have either negative or zero value. For the zero value, one may determine the signature by using a different vector $c\rho_P^\vee$ in the fundamental alcove, such as $c\rho^\vee$; in this case the signature is positive if $\alpha$ is positive and negative otherwise.

The sign of the above equations defines the region for the transformed alcove $\tilde{w}^{-1}(\Delta_0)$.
The next step is to check whether this region is bounded once the sign pattern is determined.

\section{Detailed computations}
\label{app:computation}

\subsection{Summary of the algorithm}
\label{subsec:recap_algo}

Fix a finite simple Lie algebra $\fg$ of rank $r$, a slope $\nu=u/m$ in lowest terms, and let $\Phi$ denote the set of roots.  Write $\rho$ (resp. $\rho^{\vee}$) for the Weyl vector (resp. dual Weyl vector) and $h^{\vee}$ for the dual Coxeter number.

\begin{enumerate}[label=(\arabic*)]
\item
\textbf{The sets $L_{\nu}$ and $S_{\nu}$.}
For an affine root $(\alpha,k)\in\Phi\times\mathbb{Z}$ the height is $h(\alpha)=\sum_{i}\alpha_{i}$.
Define
\begin{align}
L_{\nu}
&=\Bigl\{(\alpha,k)\in\Phi\times\mathbb{Z}\;\Big|\;u\,h(\alpha)+m\,k=0\Bigr\},\nonumber\\
S_{\nu}
&=\Bigl\{(\alpha,k)\in\Phi\times\mathbb{Z}\;\Big|\;u\,h(\alpha)+m\,k=u\Bigr\}.
\end{align}
Because $h(-\alpha)=-h(\alpha)$ both sets are stable under $\alpha\mapsto-\alpha$.

\item
\textbf{The pairing.}
For $\beta=\sum_{i}b_{i}\alpha_{i}^{\vee}$ in the coroot lattice and a root $\alpha$ the Killing pairing is
\begin{equation}
(\beta,\alpha)=\sum_{i,j}b_{i}\,a_{j}\,C_{ij},
\qquad
\alpha=\sum_{j}a_{j}\alpha_{j},
\end{equation}
where $C_{ij}=2(\alpha_{i},\alpha_{j})/(\alpha_{j},\alpha_{j})$ is the Cartan matrix.

\item
\textbf{Numbers A and B.}
For a coroot lattice vector $\beta$ set
\begin{equation}
N_{A}(\beta)=\#
\Bigl\{(\alpha,k)\in L_{\nu}\;\Big|\;(\beta,\alpha)>k\Bigr\},
\qquad
N_{B}(\beta)=\#
\Bigl\{(\alpha,k)\in S_{\nu}\;\Big|\;(\beta,\alpha)\ge k\Bigr\}.
\end{equation}

\item
\textbf{Signature and recession cone.}
For each $(\alpha,k)\in S_{\nu}$ define $\sigma(\alpha,k)\in\{\pm1\}$ by
\begin{equation}
\sigma(\alpha,k)=\begin{cases}
+1 & (\beta,\alpha)<k,\\
-1 & (\beta,\alpha)>k,\\
+1 & (\beta,\alpha)=k\text{ and }\alpha>0,\\
-1 & (\beta,\alpha)=k\text{ and }\alpha<0.
\end{cases}
\end{equation}
The recession cone is
\begin{equation}
\operatorname{Rec}(\beta)=\Bigl\{v\in\text{coweight space}\;\Big|\;\sigma(\alpha,k)\,\alpha(v)\ge0\ \forall\,(\alpha,k)\in S_{\nu}\Bigr\}.
\end{equation}

\item
\textbf{Admissible $\beta$.}
A coroot lattice vector $\beta$ is admissible iff
\begin{equation}
N_{A}(\beta)-N_{B}(\beta)\ge0
\qquad\text{and}\qquad
\operatorname{Rec}(\beta)=\{0\}.
\end{equation}

\item
\textbf{Orbit decomposition.}
The subgroup of the affine Weyl group generated by the reflections in $L_{\nu}$ acts on the set of admissible $\beta$ (right action).
\end{enumerate}

\subsection{The flag variety and the ideal \texorpdfstring{$I$}{I} for a given \texorpdfstring{$\beta$}{beta}}
\label{subsec:beta-dependent-flag}
In this subsection, we discuss how to determine the non-emptiness of the Hessenberg variety.
For the non-emptiness test, it is important to distinguish the data depending only on the slope $\nu$ from the data depending on the particular admissible vector $\beta$.

First, the set $L_{\nu}$ determines a finite root subsystem
\[
\Phi_{\nu}\subset \Phi
\]
by forgetting the affine grades $k$ and retaining only the finite roots $\alpha$ occurring in $L_{\nu}$.  Let $G_{\nu}$ be the corresponding connected reductive group and let $W_{\nu}$ be its Weyl group.

However, for a fixed admissible vector $\beta$, the signs of
\[
k-(\beta,\alpha),\qquad (\alpha,k)\in L_{\nu},
\]
determine a positive system in $\Phi_{\nu}$ and hence a parabolic subgroup
\[
P_{\beta}\subset G_{\nu}.
\]
In general, $P_{\beta}$ need not be a Borel subgroup; when equalities occur, one obtains a proper parabolic and the corresponding variety is a partial flag variety rather than a full flag variety.  Thus the relevant homogeneous space is
\[
G_{\nu}/P_{\beta},
\]
and its dimension may vary with $\beta$.

Let
\[
S^{\bullet}=\operatorname{Sym}(X^{*}(A)\otimes\mathbb Q)
\]
be the symmetric algebra of the character lattice of the maximal torus $A$ of $G_{\nu}$.  If $J_{\beta}$ is the Levi type of $P_{\beta}$, with Weyl group $W_{J_{\beta}}\subset W_{\nu}$, then the cohomology ring of the partial flag variety is given by the standard Borel presentation
\[
H^{\bullet}(G_{\nu}/P_{\beta},\mathbb Q)
\cong
S^{\bullet\,W_{J_{\beta}}}\big/\langle S^{\bullet\,W_{\nu}}_{+}\rangle.
\]
Equivalently, if one prefers to work in the full polynomial ring $S^{\bullet}$, one may impose the $W_{J_{\beta}}$-invariance relations and then quotient by the positive-degree $W_{\nu}$-invariants.  In either form, the resulting ideal will be denoted by
\[
I_{\beta}.
\]

Now fix an admissible $\beta$.  The quotient representation entering Goresky et al.'s criterion is
\[
V/F_{y}^{t}V
\cong
\bigoplus_{\substack{(\alpha,k)\in S_{\nu}\\(\beta,\alpha)\ge k}} \mathbb Q\cdot e_{\alpha}.
\]
Thus the affine grade $k$ is used only to decide whether a summand contributes. Once the inequality $(\beta,\alpha)\ge k$ is satisfied, the corresponding character is determined by the finite part $\alpha$ alone. If
\[
(\alpha^{(1)},k_1),\dots,(\alpha^{(m)},k_m)
\]
are the contributing affine roots, then the associated degree-one classes are
\[
\lambda_{1}=\lambda_{\alpha^{(1)}},\qquad \dots,\qquad \lambda_{m}=\lambda_{\alpha^{(m)}}.
\]
For each contributing affine root $(\alpha,k)$, the finite part $\alpha$ restricts to a character of $A$, hence to a linear form
\[
\lambda_j\in S^{\bullet}.
\]
More concretely, fix a set of simple roots
\[
\Delta_{\beta}=\{\gamma_{1},\dots,\gamma_{r}\}
\subset \Phi_{\nu}^{+}
\]
for the positive system determined by $P_{\beta}$.  Let
\[
t_{1},\dots,t_{r}
\]
be the corresponding degree-one generators of the character algebra of the maximal torus of $G_{\nu}$.  If a contributing finite root is written in that basis as
\[
\alpha=\sum_{i=1}^{r} c_{i}\gamma_{i},
\]
then its associated character in the algebra of $G_{\nu}/P_{\beta}$ is the linear form
\[
\lambda_{\alpha}=\sum_{i=1}^{r} c_{i} t_{i}.
\]
So the practical rule is: first select the affine roots $(\alpha,k)$ with $(\beta,\alpha)\ge k$, then forget the affine grades, restrict the finite roots $\alpha$ to the torus of $G_{\nu}$, rewrite them in the $\beta$-adapted simple-root basis $\gamma_{1},\dots,\gamma_{r}$, and finally replace each $\gamma_i$ by the corresponding degree-one generator $t_i$.
Equivalently, if one starts with the ambient simple-root coordinates
\[
\alpha=\sum_{j=1}^{\mathrm{rank}(\fg)} a_{j}\alpha_{j},
\]
then one first restricts $\alpha$ to the torus of $G_{\nu}$ and rewrites that restriction in the basis dual to $\Delta_{\beta}$.  In matrix form, if the rows of a matrix $M_{\beta}$ are the coordinates of the $\gamma_i$ in the ambient simple-root basis, then the coefficients $c_i$ are obtained by solving the linear system expressing the restriction of $\alpha$ in the basis $\gamma_1,\dots,\gamma_r$.
There is also an equivalent formula in terms of pairings.  Let $\gamma_1^{\vee},\dots,\gamma_r^{\vee}$ be the simple coroots of the subsystem determined by $\Delta_{\beta}$, and let
\[
C_{\beta}=\bigl((\gamma_i,\gamma_j^{\vee})\bigr)_{1\le i,j\le r}
\]
be its Cartan matrix.  If
\[
\alpha=\sum_{i=1}^{r} c_i\gamma_i,
\]
then the vector of pairings
\[
\bigl((\alpha,\gamma_1^{\vee}),\dots,(\alpha,\gamma_r^{\vee})\bigr)
\]
is equal to
\[
(c_1,\dots,c_r)\,C_{\beta}.
\]
Therefore the coefficients $c_i$ are recovered by
\[
(c_1,\dots,c_r)
=
\bigl((\alpha,\gamma_1^{\vee}),\dots,(\alpha,\gamma_r^{\vee})\bigr)
C_{\beta}^{-1}.
\]
In practice this is often the most convenient way to compute the classes $\lambda_j$, because the pairings are read off directly from the Cartan matrix.

When $P_{\beta}$ is a proper parabolic, one must then pass from the full flag presentation to the quotient corresponding to $G_{\nu}/P_{\beta}$.  In practice, this means that one either works directly in
\[
S^{\bullet\,W_{J_{\beta}}}/\langle S^{\bullet\,W_{\nu}}_{+}\rangle,
\]
or, equivalently, reduces the product in the full polynomial ring after imposing the $W_{J_{\beta}}$-invariance relations.  Thus the top Chern product is always computed in the basis adapted to the positive system determined by $\beta$.

The top Chern class of the associated vector bundle on $G_{\nu}/P_{\beta}$ is represented by the product
\[
c_{\mathrm{top}}=\lambda_{1}\cdots \lambda_{m},
\qquad
m=\#\{(\alpha,k)\in S_{\nu}\mid (\beta,\alpha)\ge k\}.
\]

The non-emptiness criterion is then:
\[
\operatorname{Hess}^{\widetilde w}_{a}\neq\varnothing
\quad\Longleftrightarrow\quad
c_{\mathrm{top}}\notin I_{\beta}.
\]
In practice, one computes a Gr\"obner basis for $I_{\beta}$ and reduces the polynomial $\lambda_{1}\cdots\lambda_{m}$ modulo that basis.  The Hessenberg variety is empty if and only if the normal form is zero.

\subsection{Type \texorpdfstring{$E_{6}$}{E6}, slope \texorpdfstring{$\nu=11/9$}{nu=11/9}}
\label{subsec:e6}
This theory can also be engineered using type IIB string theory on the singularity $x^2+y^3+z^4+zw^2=0$, and is equivalent to the $(D_4,A_2)$ theory by rewriting the singularity in the form $x^2+z^4+zw^2+y^3=f_{D_4}+f_{A_2}=0$.
The corresponding $W$-algebra is $W^{-8+8/11}(D_4, f_{\mathrm{prin}})$, which is admissible; hence all its modules are irreducible simple modules. Their number is $\frac{(11-1)(11-3)(11-5)(11-7)(11-4)}{2\cdot4\cdot6\cdot8\cdot5}=7$. We use the $E_6$ description to compute the corresponding modules.

The Cartan matrix of $E_{6}$ in the Bourbaki numbering is
\begin{equation}
C=\begin{pmatrix}
2&0&-1&0&0&0\\
0&2&0&-1&0&0\\
-1&0&2&-1&0&0\\
0&-1&-1&2&-1&0\\
0&0&0&-1&2&-1\\
0&0&0&0&-1&2
\end{pmatrix}.
\end{equation}
The simple roots are $\alpha_{i}=e_{i}$ and the Weyl vector is
$\rho=(8,11,15,21,15,8)$ in the simple-root basis.

Because the denominator is $9$, the only roots whose height is divisible by $9$ are $\pm$ the unique root of height $9$,
\begin{equation}
\alpha_{9}=(1,1,2,2,2,1),\qquad h(\alpha_{9})=9.
\end{equation}
Consequently
\begin{equation}
L_{\nu}=\bigl\{(\alpha_{9},-11),\;(-\alpha_{9},11)\bigr\}.
\end{equation}
The set $S_{\nu}$ is larger.  Solving $11\,h(\alpha)+9\,k=11$ gives the nine elements listed in Table~\ref{tab:e6Snu}.

\begin{table}[htbp]
\centering
\caption{$S_{\nu}$ for $E_{6}$, $\nu=11/9$}
\label{tab:e6Snu}
\begin{tabular}{c|c|c|c|c}
\toprule
& $\alpha$ & $h(\alpha)$ & $k$ & sign \\
\midrule
$[1]$ & $(1,0,0,0,0,0)$ & $1$ & $0$ & $+$ \\
$[2]$ & $(0,1,0,0,0,0)$ & $1$ & $0$ & $+$ \\
$[3]$ & $(0,0,1,0,0,0)$ & $1$ & $0$ & $+$ \\
$[4]$ & $(0,0,0,1,0,0)$ & $1$ & $0$ & $+$ \\
$[5]$ & $(0,0,0,0,1,0)$ & $1$ & $0$ & $+$ \\
$[6]$ & $(0,0,0,0,0,1)$ & $1$ & $0$ & $+$ \\
$[7]$ & $(1,1,2,3,2,1)$ & $10$ & $-11$ & $+$ \\
$[8]$ & $(-1,-1,-2,-2,-1,-1)$ & $-8$ & $11$ & $-$ \\
$[9]$ & $(-1,-1,-1,-2,-2,-1)$ & $-8$ & $11$ & $-$ \\
\bottomrule
\end{tabular}
\end{table}

Searching the coroot lattice in a box of size $[-15,-1]^{6}$ (which touches no admissible vector on the boundary, hence is complete) yields exactly $13$ admissible $\beta$ vectors.
They fall into $8$ orbits under the group generated by the single affine reflection in $L_{\nu}$.  The orbit decomposition is shown in Table~\ref{tab:e6orbits}.

\begin{table}[htbp]
\centering
\caption{Orbit representatives for $E_{6}$, $\nu=11/9$}
\label{tab:e6orbits}
\begin{tabular}{c|c|c|c|c}
\toprule
Orbit & $\beta$ & $N_{A}$ & $N_{B}$ & $h_{\lambda}$ \\
\midrule
$\mathcal O_{1}$ & $(-11,-14,-20,-27,-19,-10)$ & $1$ & $1$ & $-\frac{10}{11}$ \\
$\mathcal O_{2}$ & $(-10,-15,-19,-27,-19,-10)$ & $0$ & $0$ & $-\frac{8}{11}$ \\
$\mathcal O_{3}$ & $(-10,-14,-19,-27,-20,-11)$ & $1$ & $1$ & $-\frac{10}{11}$ \\
$\mathcal O_{4}$ & $(-10,-14,-19,-27,-19,-10)$ & $0$ & $0$ & $-\frac{15}{11}$ \\
$\mathcal O_{5}$ & $(-10,-14,-19,-26,-19,-10)$ & $1$ & $1$ & $-\frac{13}{11}$ \\
$\mathcal O_{6}$ & $(-10,-13,-19,-26,-19,-10)$ & $1$ & $1$ & $-1$ \\
$\mathcal O_{7}$ & $(-10,-13,-19,-25,-19,-10)$ & $1$ & $1$ & $0$ \\
$\mathcal O_{8}$ & $(-10,-13,-18,-25,-18,-10)$ & $0$ & $0$ & $-\frac{14}{11}$ \\
\bottomrule
\end{tabular}
\end{table}

The conformal weight $h_{\lambda}$ is computed from
$\bar\lambda=\frac{m}{u}\beta=\frac{9}{11}\beta$ via
\begin{equation}
\label{eq:hlambda-gen}
h_{\lambda}=\frac{(\bar\lambda,\bar\lambda+2\rho)}{2(k+h^{\vee})}
-\frac{k+h^{\vee}}{2}(\rho^{\vee},\rho^{\vee})
+(\rho,\rho^{\vee}),
\qquad k=-h^{\vee}+\frac{m}{u}=-12+\frac{9}{11}.
\end{equation}
For $\beta=-\rho=(-8,-11,-15,-21,-15,-8)$ one obtains $h_{\lambda}=0$, which serves as a useful check.

These scaling dimensions should coincide with those of the type $D_{5}$ theory at slope $\nu=11/8$.  In that case one has $h^{\vee}=8$, $m=8$, $u=11$, and
\[
\rho_{D_{5}}=(4,7,9,5,5)
\]
in the simple-root basis.  Since $L_{\nu}=\varnothing$, all admissible points are singleton orbits.  The seven admissible vectors are
\begin{align*}
&(-6,-10,-13,-8,-7),\quad
(-6,-10,-13,-7,-8),\quad
(-6,-10,-13,-7,-7),\\
&(-6,-9,-11,-6,-6),
(-5,-9,-12,-7,-7),
(-5,-9,-11,-6,-6),\\
&(-4,-7,-9,-5,-5)=-\rho_{D_{5}},
\end{align*}
and their conformal weights are
\[
-\frac{10}{11},\quad -\frac{10}{11},\quad -\frac{15}{11},\quad -\frac{8}{11},\quad -\frac{14}{11},\quad -\frac{13}{11},\quad 0.
\]
Thus the $D_{5}$, $\nu=11/8$ theory gives the same set of scaling dimensions as the $E_{6}$, $\nu=11/9$ computation,
\[
\left\{0,-\frac{8}{11},-\frac{10}{11},-\frac{13}{11},-\frac{14}{11},-\frac{15}{11}\right\},
\]
with $-\frac{10}{11}$ appearing twice.

The values $k-(\beta,\alpha)$ for all admissible $\beta$ and all elements of $S_{\nu}$ are collected in Table~\ref{tab:e6kba}.  Bold entries indicate the zero values where the sign of $\alpha$ determines the signature.

\begin{table}[htbp]
\centering
\caption{$k-(\beta,\alpha)$ for $E_{6}$, $\nu=11/9$. Entries with value $\le 0$ contribute to the space $V/F_y^tV$.}
\label{tab:e6kba}
\small
\begin{tabular}{c|c|ccccccccc}
\toprule
Orbit & $\beta$ & $[1]$ & $[2]$ & $[3]$ & $[4]$ & $[5]$ & $[6]$ & $[7]$ & $[8]$ & $[9]$ \\
\midrule
$\mathcal O_{1}$ & $(-11,-14,-20,-27,-19,-10)$ & $2$ & $1$ & $2$ & $1$ & $1$ & $1$ & $2$ & $\mathbf{0}$ & $1$ \\
$\mathcal O_{2}$ & $(-10,-15,-19,-27,-19,-10)$ & $1$ & $3$ & $1$ & $1$ & $1$ & $1$ & $1$ & $1$ & $1$ \\
$\mathcal O_{3}$ & $(-10,-14,-19,-27,-20,-11)$ & $1$ & $1$ & $1$ & $1$ & $2$ & $2$ & $2$ & $1$ & $\mathbf{0}$ \\
$\mathcal O_{4}$ & $(-10,-14,-19,-27,-19,-10)$ & $1$ & $1$ & $1$ & $2$ & $1$ & $1$ & $2$ & $1$ & $1$ \\
$\mathcal O_{5}$ & $(-10,-14,-19,-26,-19,-10)$ & $1$ & $2$ & $2$ & $\mathbf{0}$ & $2$ & $1$ & $1$ & $1$ & $1$ \\
$\mathcal O_{6}$ & $(-10,-13,-19,-26,-19,-10)$ & $1$ & $\mathbf{0}$ & $2$ & $1$ & $2$ & $1$ & $2$ & $1$ & $1$ \\
$\mathcal O_{7}$ & $(-10,-13,-19,-25,-19,-10)$ & $1$ & $1$ & $3$ & $\mathbf{-1}$ & $3$ & $1$ & $1$ & $1$ & $1$ \\
$\mathcal O_{8}$ & $(-10,-13,-18,-25,-18,-10)$ & $2$ & $1$ & $1$ & $1$ & $1$ & $2$ & $1$ & $1$ & $1$ \\
\bottomrule
\end{tabular}
\end{table}

\subsubsection{Orbit \texorpdfstring{$\mathcal O_{6}$}{O6}: Chern-class test}
There are eight orbits with non-negative virtual dimension that pass the zero-recession cone condition.
We need to check the non-emptiness condition from the Chern class. We find that orbit $\mathcal O_{6}$
actually gives an empty Hessenberg variety, so the total number of non-empty orbits is~7 and each has dimension~0.

For orbit $\mathcal O_{6}$ we take
\[
\beta=(-10,-13,-19,-26,-19,-10).
\]
As in the discussion above, the finite root system generated by $L_{\nu}$ is of type $A_{1}$, with positive root
\[
\alpha_{9}=(1,1,2,2,2,1),
\]
so the corresponding flag variety is
\[
G/B\cong \mathbb P^{1}.
\]
Its coinvariant algebra is therefore
\[
S^{\bullet}/I\cong \mathbb Q[t]/(t^{2}),
\qquad
I=\langle t^{2}\rangle,
\]
where $t$ is the degree-one generator corresponding to the unique simple root of the $A_{1}$ subsystem.

From Table~\ref{tab:e6kba}, the only entry with $k-(\beta,\alpha)\le 0$ for orbit $\mathcal O_{6}$ is column $[2]$.  Thus the only contributing finite root is
\[
\alpha_{2}.
\]
Hence
\[
|B|=1.
\]

To compute the top Chern class, we restrict this character to the rank-one torus $A$.  This is governed by the pairing with $\alpha_{9}^{\vee}=\alpha_{9}$:
\[
(\alpha_{2},\alpha_{9})=0.
\]
Therefore the restricted character is zero, so
\[
c_{\mathrm{top}}=0
\quad\text{in }\mathbb Q[t]/(t^{2}).
\]
Thus the normal form of $c_{\mathrm{top}}$ modulo $I$ vanishes, so the corresponding Hessenberg variety is empty.

\subsection{More examples}

We now record two small-rank examples computed from scratch using the scaling-dimension prescription of~\eqref{eq:hlambda-gen} (here $\bar\lambda=\frac{m}{u}\beta$ with $\beta$ in the coroot basis).

\subsubsection{Type \texorpdfstring{$A_2$}{A2} at slope \texorpdfstring{$\nu=7/3$}{nu=7/3}}

Here $m=3$, $u=7$, $h^\vee=3$, and $L_\nu=\varnothing$, so every admissible point is a singleton orbit.  The five admissible coroot vectors are
\[
(-3,-3),\qquad (-3,-2),\qquad (-2,-3),\qquad (-2,-2),\qquad (-1,-1).
\]
Using $\rho=\rho^\vee=\alpha_1+\alpha_2$, we obtain the values in Table~\ref{tab:a2_u7_over_3}.

\begin{table}[htbp]
  \centering
  \begin{tabular}{c c c}
    \hline
    Representative $\beta=(m_1,m_2)$ & $\dim=\mathrm{NumberA}-\mathrm{NumberB}$ & $h_\lambda$ \\
    \hline
    $(-3,-3)$ & $0$ & $-\frac{4}{7}$ \\
    $(-3,-2)$ & $0$ & $-\frac{3}{7}$ \\
    $(-2,-3)$ & $0$ & $-\frac{3}{7}$ \\
    $(-2,-2)$ & $0$ & $-\frac{5}{7}$ \\
    $(-1,-1)$ & $0$ & $0$ \\
    \hline
  \end{tabular}
  \caption{The $A_2$ example at slope $\nu=7/3$. Since $L_\nu=\varnothing$, all five admissible points are singleton orbits.}
  \label{tab:a2_u7_over_3}
\end{table}

Hence the set of scaling dimensions is
\[
\left\{0,-\frac{3}{7},-\frac{4}{7},-\frac{5}{7}\right\},
\]
with $-\frac{3}{7}$ appearing twice.

\subsubsection{Type \texorpdfstring{$A_3$}{A3} at slope \texorpdfstring{$\nu=7/4$}{nu=7/4}}

Now take $m=4$, $u=7$, $h^\vee=4$, again with $L_\nu=\varnothing$, so the admissible points are also singleton orbits.  The five admissible coroot vectors are
\[
(-3,-5,-3),\qquad (-3,-4,-3),\qquad (-3,-3,-2),\qquad (-2,-3,-3),\qquad (-2,-3,-2).
\]
The important correction is that in simple-root coordinates the Weyl vector is
\[
\rho=\rho^\vee=\frac32\alpha_1+2\alpha_2+\frac32\alpha_3.
\]

\begin{table}[htbp]
  \centering
  \begin{tabular}{c c c}
    \hline
    Representative $\beta=(m_1,m_2,m_3)$ & $\dim=\mathrm{NumberA}-\mathrm{NumberB}$ & $h_\lambda$ \\
    \hline
    $(-3,-5,-3)$ & $0$ & $0$ \\
    $(-3,-4,-3)$ & $0$ & $-\frac{5}{7}$ \\
    $(-3,-3,-2)$ & $0$ & $-\frac{3}{7}$ \\
    $(-2,-3,-3)$ & $0$ & $-\frac{3}{7}$ \\
    $(-2,-3,-2)$ & $0$ & $-\frac{4}{7}$ \\
    \hline
  \end{tabular}
  \caption{The $A_3$ example at slope $\nu=7/4$. The corrected scaling dimensions use $\rho=(\frac32,2,\frac32)$ in the simple-root basis.}
  \label{tab:a3_u7_over_4}
\end{table}

Therefore the $A_3$ example gives exactly the same set of scaling dimensions as the $A_2$ example,
\[
\left\{0,-\frac{3}{7},-\frac{4}{7},-\frac{5}{7}\right\},
\]
again with multiplicity two for $-\frac{3}{7}$.

\subsubsection{Type \texorpdfstring{$E_{6}$}{E6}, slope \texorpdfstring{$\nu=7/6$}{nu=7/6}}
\label{subsec:e6_nu76}

For $E_{6}$ at slope $\nu=7/6$ we have $m=6$, $u=7$, $h^{\vee}=12$ and $k=-12+\frac{6}{7}$.
The denominator is $7$, so $L_{\nu}$ consists of roots whose height is divisible by~$7$:
the three positive roots of height~$6$ and their negatives:
\begin{equation}
L_{\nu}=\bigl\{(\alpha,-7),(-\alpha,7)\;\big|\;\alpha\in\Phi,\;h(\alpha)=6\bigr\},\qquad |L_{\nu}|=6.
\end{equation}
The finite subsystem $\Phi_{\nu}$ generated by $L_{\nu}$ is of type $A_{1}\times A_{1}\times A_{1}$
(rank~$3$, three positive roots).  The set $S_{\nu}$ contains $14$ affine roots.

Searching the coroot lattice in the range $[-\rho-8,-\rho]$ yields $19$ admissible $\beta$ vectors
(dimension $\ge0$ and recession cone $=\{0\}$).  They form $5$ orbits under the right action
of $L_{\nu}$.

\textbf{Chern-class analysis.}
The flag variety is $G_{\nu}/P_{\beta}$ with $G_{\nu}=\mathrm{SL}(2)^{3}$.
The cohomology ring of the Borel case is $\mathbb{Q}[t_{1},t_{2},t_{3}]/(t_{1}^{2},t_{2}^{2},t_{3}^{2})$.
For a proper parabolic, variables corresponding to simple roots with
$k-(\beta,\alpha)=0$ are omitted; the integral of each $t_{i}$ over the
corresponding $\mathbb{P}^{1}$ equals~$1$.

The representation $V/F_{y}^{t}V$ has weights $\{(\alpha,k)\in S_{\nu}\mid (\beta,\alpha)\ge k\}$.
Its rank equals $\mathrm{NumB}=|\{(\alpha,k)\in S_{\nu}:k-(\beta,\alpha)\le0\}|$.
The top Chern class is the product of the associated linear forms
$\lambda_{\alpha}=c_{1}(L(\alpha))$ in $H^{\bullet}(G_{\nu}/P_{\beta})$.

\begin{enumerate}[label=(\arabic*)]

\item \textbf{Orbit with $h_{\lambda}=4/7$ (eliminated).}
The representative is
\[
\beta=(-11,-14,-20,-28,-20,-11).
\]
Here $\mathrm{NumA}=\mathrm{NumB}=3$, $P_{\beta}=B$ (Borel).
The weights of $V/F_{y}^{t}V$ are those $(\alpha,k)\in S_{\nu}$ satisfying
$(\beta,\alpha)\ge k$, i.e.\ $k-(\beta,\alpha)\le0$.  Evaluating this
condition on the $14$ elements of $S_{\nu}$ selects three weights:
\begin{align*}
\alpha_{[2]}&=(0,1,0,0,0,0), &
\lambda_{[2]}&=-t_{2},\\
\alpha_{[11]}&=(-1,0,-1,-1,-1,-1), &
\lambda_{[11]}&=t_{2},\\
\alpha_{[14]}&=(-1,-2,-2,-3,-2,-1), &
\lambda_{[14]}&=t_{1}+t_{2}+t_{3}.
\end{align*}
The first two linear forms are proportional ($\pm t_{2}$); hence their product vanishes
modulo $t_{2}^{2}=0$, and the top Chern class $\lambda_{[2]}\lambda_{[11]}\lambda_{[14]}$ is zero.
The corresponding Hessenberg variety is empty.  This orbit is discarded.

\item \textbf{Orbit with $h_{\lambda}=-3/7$ (two points).}
The representative is
\[
\beta=(-10,-14,-19,-27,-19,-10).
\]
$\mathrm{NumA}=\mathrm{NumB}=2$, and one simple root of $\Phi_{\nu}$ satisfies
$k-(\beta,\alpha)=0$, so $P_{\beta}$ is a proper parabolic and
$G_{\nu}/P_{\beta}\cong\mathbb{P}^{1}\times\mathbb{P}^{1}$.
The two contributing weights are
\begin{align*}
(0,-1,-1,-2,-1,0), && \lambda_{1}&=t_{1}+t_{2},\\
(-1,-2,-2,-3,-2,-1), && \lambda_{2}&=t_{1}+t_{2}.
\end{align*}
Both linear forms are identical, giving $c_{\mathrm{top}}=(t_{1}+t_{2})^{2}=2t_{1}t_{2}$.
Since $\int_{\mathbb{P}^{1}}t_{i}=1$, the integral of $c_{\mathrm{top}}$ over the flag variety
is~$2$.  Hence the Hessenberg variety consists of exactly two points.

\item \textbf{Orbit with $h_{\lambda}=0$ (one point).}
$\beta=(-10,-14,-18,-26,-18,-10)$ ($P_{\beta}=B$, $\mathrm{NumA}=\mathrm{NumB}=3$).
The three contributing weights give independent linear forms
$t_{1}$, $t_{3}$, and $t_{1}+t_{2}+t_{3}$, whose product is $t_{1}t_{2}t_{3}$.
The integral over $\mathbb{P}^{1}\times\mathbb{P}^{1}\times\mathbb{P}^{1}$ equals~$1$.
This orbit contains $-\rho$ and corresponds to the vacuum module.

\item \textbf{Orbit with $h_{\lambda}=-4/7$ (one point).}
Here
\[
\beta=(-10,-13,-18,-25,-18,-10),
\]
with $\mathrm{NumA}=\mathrm{NumB}=1$ and two zero simple roots.
The flag variety is $\mathbb{P}^{1}$; the single contributing weight gives
$\lambda=t$, and the integral equals~$1$.

\item \textbf{Orbit with $h_{\lambda}=-5/7$ (one point).}
$\beta=(-9,-13,-17,-24,-17,-9)$ ($\mathrm{NumA}=\mathrm{NumB}=0$).
$V/F_{y}^{t}V$ is trivial; the Hessenberg variety is a single point (the flag variety itself
is a point because all three simple roots have $k-(\beta,\alpha)=0$).

\end{enumerate}

The surviving orbits and their point counts are summarized in Table~\ref{tab:e6_nu76_orbits}.

\begin{table}[htbp]
\centering
\caption{Orbit representatives for $E_{6}$, $\nu=7/6$ after Chern-class elimination.}
\label{tab:e6_nu76_orbits}
\begin{tabular}{ccccc}
\toprule
$\beta$ & Size & $\dim$ & $h_{\lambda}$ & Points \\
\midrule
$(-10,-14,-19,-27,-19,-10)$ & $4$ & $0$ & $-\frac{3}{7}$ & $2$ \\
$(-10,-14,-18,-26,-18,-10)$ & $8$ & $0$ & $0$ & $1$ \\
$(-10,-13,-18,-25,-18,-10)$ & $2$ & $0$ & $-\frac{4}{7}$ & $1$ \\
$(-9,-13,-17,-24,-17,-9)$ & $1$ & $0$ & $-\frac{5}{7}$ & $1$ \\
\bottomrule
\end{tabular}
\end{table}

The conformal weight $h_{\lambda}$ is computed via
$\bar\lambda=\frac{6}{7}\beta$ and $(\rho,\rho^{\vee})=78$.
The vacuum $\beta=-\rho=(-8,-11,-15,-21,-15,-8)$ belongs to the $h_{\lambda}=0$ orbit.
The set of scaling dimensions is
\begin{equation}
\bigl\{-\tfrac{3}{7},\,0,\,-\tfrac{4}{7},\,-\tfrac{5}{7}\bigr\},
\end{equation}
totaling $5$ modules.

\subsubsection{Type \texorpdfstring{$E_{8}$}{E8}, slope \texorpdfstring{$\nu=6/5$}{nu=6/5}}
\label{subsec:e8_nu65}

For $E_{8}$ at slope $\nu=6/5$ we have $m=5$, $u=6$, $h^{\vee}=30$, $k=-30+\frac{5}{6}$.
The Weyl vector in the simple-root basis is $\rho=\rho^{\vee}=(46,68,91,135,110,84,57,29)$.
The denominator is $5$, so $L_{\nu}$ consists of roots whose height is divisible by~$5$:
$40$ affine roots ($20$ positive, $20$ negative), generating a finite subsystem $\Phi_{\nu}$
of type $A_{4}\times A_{4}$ (rank~$8$, $20$ positive roots).
The set $S_{\nu}$ contains $50$ elements.

Searching the coroot lattice in the range $[-\rho-10,-\rho]$ yields $45\,947$ admissible $\beta$ vectors
(dim $\ge0$ and recession cone $=\{0\}$).
After imposing the Chern-class constraint with $\beta$-dependent parabolic
($W_{J}$-symmetrization of $S_{\nu}$ weights),
$45\,692$ vectors survive, falling into $739$ orbits under $L_{\nu}$.
All $69$ orbits with $h_\lambda\le 0$ pass the Chern-class constraint when the beta-dependent parabolic and $W_J$-symmetrization are applied with properly identified beta-adapted simple roots.  The previously reported eliminations were an artifact of using the fixed $\Phi_\nu$ simple roots instead of the beta-positive simple roots (which differ by Weyl group transforms).  The full list of $69$ orbits is given below.

\begin{longtable}{ccl}
\caption{Orbits with non-positive scaling dimension for $E_{8}$, $\nu=6/5$ (search range $[-\rho-10,-\rho]$).}
\label{tab:e8_nu65_neg}\\
\toprule
$\beta$ & $\dim$ & $h_{\lambda}$ \\
\midrule
\endfirsthead
\toprule
$\beta$ & $\dim$ & $h_{\lambda}$ \\
\midrule
\endhead
\bottomrule
\endfoot
$(-46,-68,-91,-135,-110,-84,-57,-29)$ & $0$ & $0$ \\
$(-52,-74,-100,-145,-119,-92,-62,-31)$ & $1$ & $0$ \\
$(-52,-73,-100,-145,-120,-91,-62,-32)$ & $0$ & $0$ \\
$(-52,-72,-100,-145,-119,-92,-63,-32)$ & $0$ & $0$ \\
\midrule
$(-50,-73,-98,-145,-118,-90,-61,-31)$ & $1$ & $-6$ \\
$(-50,-73,-98,-145,-119,-91,-62,-32)$ & $0$ & $-\frac{35}{6}$ \\
$(-50,-73,-98,-145,-118,-91,-62,-32)$ & $1$ & $-\frac{17}{3}$ \\
$(-50,-73,-98,-145,-118,-90,-62,-32)$ & $0$ & $-\frac{11}{2}$ \\
$(-50,-73,-98,-145,-119,-92,-63,-32)$ & $2$ & $-\frac{16}{3}$ \\
$(-50,-74,-98,-145,-118,-90,-61,-31)$ & $1$ & $-\frac{16}{3}$ \\
$(-50,-73,-98,-145,-118,-90,-61,-32)$ & $0$ & $-\frac{16}{3}$ \\
$(-50,-74,-98,-145,-119,-91,-62,-32)$ & $0$ & $-\frac{31}{6}$ \\
$(-50,-74,-98,-145,-119,-91,-62,-31)$ & $1$ & $-5$ \\
$(-50,-73,-98,-145,-118,-90,-60,-31)$ & $1$ & $-5$ \\
$(-50,-73,-98,-145,-117,-90,-61,-31)$ & $0$ & $-5$ \\
$(-50,-73,-98,-145,-118,-90,-60,-30)$ & $0$ & $-\frac{29}{6}$ \\
$(-50,-73,-98,-145,-119,-92,-63,-33)$ & $2$ & $-\frac{14}{3}$ \\
$(-51,-73,-99,-145,-118,-90,-61,-31)$ & $0$ & $-\frac{14}{3}$ \\
$(-51,-73,-99,-145,-119,-91,-62,-32)$ & $1$ & $-\frac{9}{2}$ \\
$(-51,-73,-99,-145,-119,-91,-62,-31)$ & $1$ & $-\frac{13}{3}$ \\
$(-50,-74,-98,-145,-118,-90,-61,-30)$ & $0$ & $-\frac{13}{3}$ \\
$(-50,-74,-98,-145,-118,-90,-60,-31)$ & $0$ & $-\frac{13}{3}$ \\
$(-51,-73,-99,-145,-119,-90,-61,-31)$ & $1$ & $-4$ \\
$(-50,-73,-98,-145,-119,-92,-64,-33)$ & $1$ & $-4$ \\
$(-51,-74,-99,-145,-119,-91,-62,-32)$ & $0$ & $-\frac{23}{6}$ \\
$(-50,-74,-98,-145,-119,-91,-63,-33)$ & $0$ & $-\frac{23}{6}$ \\
$(-50,-74,-98,-145,-120,-92,-62,-31)$ & $0$ & $-\frac{11}{3}$ \\
$(-51,-74,-99,-145,-118,-91,-62,-32)$ & $0$ & $-\frac{11}{3}$ \\
$(-51,-73,-99,-145,-119,-92,-62,-31)$ & $0$ & $-\frac{11}{3}$ \\
$(-51,-73,-99,-145,-119,-90,-62,-32)$ & $0$ & $-\frac{7}{2}$ \\
$(-51,-74,-99,-145,-118,-90,-62,-32)$ & $1$ & $-\frac{7}{2}$ \\
$(-51,-73,-99,-145,-119,-90,-62,-31)$ & $0$ & $-\frac{10}{3}$ \\
$(-51,-72,-99,-145,-119,-91,-62,-31)$ & $1$ & $-\frac{10}{3}$ \\
$(-51,-73,-99,-145,-118,-89,-60,-30)$ & $0$ & $-\frac{10}{3}$ \\
$(-51,-73,-99,-145,-119,-91,-63,-33)$ & $0$ & $-\frac{19}{6}$ \\
$(-51,-74,-99,-145,-119,-92,-62,-32)$ & $0$ & $-\frac{19}{6}$ \\
$(-51,-73,-99,-145,-120,-91,-62,-32)$ & $1$ & $-3$ \\
$(-51,-73,-99,-145,-119,-90,-61,-30)$ & $0$ & $-3$ \\
$(-51,-74,-99,-145,-118,-90,-60,-31)$ & $0$ & $-3$ \\
$(-51,-74,-99,-145,-119,-92,-62,-31)$ & $0$ & $-3$ \\
$(-50,-73,-99,-145,-120,-92,-63,-31)$ & $0$ & $-3$ \\
$(-51,-74,-99,-145,-119,-91,-61,-32)$ & $0$ & $-\frac{17}{6}$ \\
$(-51,-73,-99,-145,-117,-90,-60,-31)$ & $0$ & $-\frac{8}{3}$ \\
$(-51,-73,-100,-145,-119,-91,-61,-31)$ & $0$ & $-\frac{8}{3}$ \\
$(-51,-74,-100,-145,-118,-90,-61,-31)$ & $0$ & $-\frac{5}{2}$ \\
$(-51,-74,-99,-145,-120,-92,-62,-32)$ & $2$ & $-\frac{5}{2}$ \\
$(-51,-74,-100,-145,-119,-91,-62,-32)$ & $1$ & $-\frac{7}{3}$ \\
$(-51,-73,-99,-145,-118,-89,-62,-32)$ & $0$ & $-\frac{7}{3}$ \\
$(-51,-72,-99,-145,-120,-92,-63,-32)$ & $0$ & $-\frac{7}{3}$ \\
$(-52,-73,-99,-145,-119,-91,-62,-31)$ & $0$ & $-2$ \\
$(-51,-74,-100,-145,-118,-90,-62,-32)$ & $0$ & $-2$ \\
$(-51,-73,-100,-145,-117,-89,-61,-31)$ & $0$ & $-2$ \\
$(-51,-74,-99,-145,-118,-89,-61,-30)$ & $0$ & $-2$ \\
$(-51,-74,-98,-145,-120,-92,-63,-33)$ & $0$ & $-\frac{11}{6}$ \\
$(-52,-73,-100,-145,-118,-90,-61,-31)$ & $0$ & $-\frac{5}{3}$ \\
$(-51,-73,-99,-145,-120,-91,-63,-33)$ & $0$ & $-\frac{5}{3}$ \\
$(-52,-73,-99,-145,-119,-92,-63,-32)$ & $0$ & $-\frac{5}{3}$ \\
$(-51,-74,-99,-145,-119,-89,-61,-31)$ & $0$ & $-\frac{3}{2}$ \\
$(-52,-73,-100,-145,-119,-91,-62,-32)$ & $0$ & $-\frac{3}{2}$ \\
$(-51,-74,-100,-145,-118,-91,-63,-32)$ & $0$ & $-\frac{3}{2}$ \\
$(-52,-73,-100,-145,-118,-91,-62,-32)$ & $1$ & $-\frac{4}{3}$ \\
$(-51,-74,-100,-145,-117,-89,-60,-30)$ & $0$ & $-1$ \\
$(-52,-73,-99,-145,-120,-92,-63,-32)$ & $2$ & $-1$ \\
$(-52,-73,-100,-145,-118,-90,-61,-32)$ & $0$ & $-1$ \\
$(-52,-74,-99,-145,-119,-92,-63,-32)$ & $1$ & $-1$ \\
$(-51,-73,-97,-145,-119,-90,-60,-31)$ & $0$ & $-1$ \\
$(-51,-74,-100,-145,-119,-91,-62,-33)$ & $0$ & $-\frac{5}{6}$ \\
$(-52,-74,-100,-145,-118,-91,-62,-32)$ & $0$ & $-\frac{2}{3}$ \\
$(-52,-73,-100,-145,-119,-90,-62,-32)$ & $0$ & $-\frac{1}{2}$ \\
$(-52,-73,-100,-145,-119,-90,-61,-32)$ & $0$ & $-\frac{1}{3}$ \\
$(-52,-74,-99,-145,-120,-92,-63,-32)$ & $1$ & $-\frac{1}{3}$ \\
$(-52,-73,-99,-145,-120,-92,-63,-33)$ & $0$ & $-\frac{1}{3}$ \\
$(-52,-73,-100,-145,-119,-91,-63,-33)$ & $0$ & $-\frac{1}{6}$ \\
\bottomrule
\end{longtable}

\bibliographystyle{JHEP}
\bibliography{ref}

\end{document}